\documentclass[journal,doublecolumn]{IEEEtran}
\usepackage{stfloats}
\usepackage{cite}
\usepackage{amsmath,amssymb,amsfonts}
\usepackage{graphicx}
\usepackage{subfigure}
\usepackage{textcomp}
\usepackage{caption}
\usepackage{xcolor}
\def\BibTeX{{\rm B\kern-.05em{\sc i\kern-.025em b}\kern-.08em
    T\kern-.1667em\lower.7ex\hbox{E}\kern-.125emX}}
\usepackage{algorithm}  
\usepackage{algorithmic}
\usepackage{amsmath} 
\usepackage{adjustbox}

\usepackage{caption}
\usepackage{flushend}

\begin{document}

\title{Improving Information Freshness via Backbone-Assisted Cooperative Access Points\\
}

\author{Haoyuan~Pan,~\IEEEmembership{Member,~IEEE,}~Yu~Zhou,~Tse-Tin~Chan,~\IEEEmembership{Member,~IEEE,}\\~Ming~Tang,~\IEEEmembership{Member,~IEEE,}~Jianqiang~Li,~\IEEEmembership{Member,~IEEE,}~Zhihua Du%

\thanks{H. Pan, Y. Zhou, J. Li, and Z. Du are with the College of Computer Science and Software Engineering, Shenzhen University, Shenzhen, China (e-mails: {hypan@szu.edu.cn}, {zhouyu2020@email.szu.edu.cn}, {lijq@szu.edu.cn}, {duzh@szu.edu.cn}). }
\thanks{T.-T.~Chan is with the Department of Mathematics and Information Technology, The Education University of Hong Kong, Hong Kong SAR, China (e-mail: {tsetinchan@eduhk.hk}).}
\thanks{M. Tang is with the Department of Computer Science and Engineering in Southern University of Science and Technology, Shenzhen, China (e-mail: {tangm3@sustech.edu.cn}).}
}

\maketitle

\begin{abstract}
Information freshness, characterized by age of information (AoI), is important for sensor applications involving timely status updates. In many cases, the wireless signals from one sensor can be received by multiple access points (APs). This paper investigates the average AoI for cooperative APs, in which they can share information through a wired backbone network. We first study a basic backbone-assisted COoperative AP (Co-AP) system where APs share only decoded packets. Experimental results on software-defined radios (SDR) indicate that Co-AP significantly improves the average AoI performance over a single-AP system. Next, we investigate an improved Co-AP system, called Soft-Co-AP. In addition to sharing decoded packets, Soft-Co-AP shares and collects soft information of packets that the APs fail to decode for further joint decoding. A critical issue in Soft-Co-AP is determining the number of quantization bits that represent the soft information (each soft bit) shared over the backbone. While more quantization bits per soft bit improves the joint decoding performance, it leads to higher backbone delay. We experimentally study the average AoI of Soft-Co-AP by evaluating the tradeoff between the backbone delay and the number of quantization bits. SDR experiments show that when the number of sensors is large, Soft-Co-AP further reduces the average AoI by $12$\% compared with Co-AP. Interestingly, good average AoI performance is usually achieved when the number of quantization bits per soft bit is neither too large nor too small.
\end{abstract}


\section{Introduction}
\begin{figure}[t]
\centerline{\includegraphics[trim=0 0 0 00,width=0.5\textwidth]{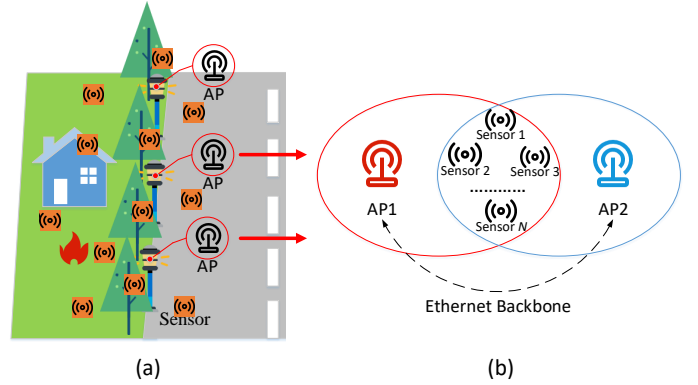}}
\caption{(a) Status update in smart cities: a large number of sensors are distributed on the street, and they send measurements to access points (AP) installed on the lamp posts. (b) A simplified backbone-assisted cooperative AP scenario: there are \emph{N} sensors located in the overlapping coverage area of two APs, namely AP1 and AP2. An Ethernet backbone connects the two APs, allowing APs to exchange information.}
\label{system_model}
\end{figure}

The Internet of Things (IoT) is envisioned to be a fundamental enabler for future wireless communication networks by interconnecting the physical world into computer networks \cite{A. Zanella}. In recent years, the explosive growth of IoT devices with sensing, communication, and data analytics capabilities has brought new applications requiring timely status updates \cite{M. A. Abd-Elmagid,D. Ciuonzo,M. A. Al-Jarrah,X. Cheng}. For example, in future smart cities shown in  Fig.~\ref{system_model}(a), a large number of sensors distributed on the street sample and send measurements of physical characteristics, such as temperature, pollution index, and traffic flow, to access points (AP) installed on the lamp posts. In such a smart city scenario, the freshness of sensed data is of paramount importance. For example, in future vehicle-to-everything (V2X) networks, fresh traffic flow monitoring information is critical to enhance mutual awareness of the surroundings and to reduce the risk of road accidents \cite{H. Zhou}.

Age of information (AoI), a fundamental metric to quantify information freshness, was first proposed in \cite{S. Kaul}. It measures the time elapsed since the generation time of the latest status update received at the receiver \cite{A. Kosta}. More specifically, if at time $t$, the latest status update received at the receiver was an update packet generated at time $t-\tau$ at the transmitter (say, a sensor), then the \emph{instantaneous} AoI of the sensor is $\tau$. Since its introduction in \cite{S. Kaul}, AoI has attracted considerable research interest, and the \emph{average} AoI is the most commonly used metric for measuring information freshness, i.e., the time average of instantaneous AoI \cite{A. Kosta, M. A. Abd-Elmagid}. Prior works revealed that since AoI captures both the generation time of an update packet and its delay through the network, optimizing the average AoI is usually different from optimizing packet delay \cite{A. Kosta}. In addition, various techniques were investigated to improve the average AoI, such as channel coding \cite{X. Chen, H. Pan2}, advanced multiple access schemes \cite{H. Pan1, Munari,Grybosi} (see Section \ref{sec:relatedwork} for more details). 

In Fig.~\ref{system_model}(a), due to the dense deployment of APs and sensors, multiple APs can receive packets from the same sensor. As a first step, we focus on two neighboring APs among many APs for cooperation. To be specific, let us consider a simplified scenario shown in Fig.~\ref{system_model}(b), in which there are $N$ sensors located in the overlapping coverage area of two APs, namely AP1 and AP2. This means that both AP1 and AP2 can hear these $N$ sensors. Without loss of generality, suppose that AP1 and AP2 serve as the primary AP and the secondary AP, respectively, and the update packets from sensors are destined for the primary AP. Suppose AP1 fails to receive an update packet due to wireless impairments (such as shadowing and fading), but AP2 succeeds. In that case, AP2 can forward the received packet to AP1 through the backbone network infrastructure interconnecting AP1 and AP2 in smart cities. Often the backbone connection between the two APs is wired. This packet forwarding mechanism over the backbone network essentially increases the packet reception probability through spatial diversity: it only requires that at least one of the APs receive the packet. In other words, a packet that reaches more than one AP can have a higher chance of being received, thus potentially improving information freshness. 

In this paper, we investigate Ethernet-backbone-assisted cooperative APs to reduce the average AoI of sensor networks with timely status update requirements. We first study a basic backbone-assisted cooperative AP system, referred to as the Co-AP system, in which only the decoded packets are forwarded through the Ethernet backbone. We consider multiple sensors sending update packets in a time division multiple access (TDMA) manner as a simple implementation. Thanks to the backbone connection, AP1 can receive sensors' update packets from AP2 when packets fail to be decoded at AP1 but are decoded successfully at AP2. We note that an update packet of a sensor usually reaches AP1 via the backbone before the next update opportunity of the same sensor (i.e., the next TDMA slot allocated to this sensor). Section \ref{sec:decoded_packcet} shows that this phenomenon effectively improves the information freshness. Our experiments on software-defined radio (SDR) show that compared to the traditional system with non-cooperative APs (i.e., the single-AP system), Co-AP reduces
the average AoI by around $50\%$ when the number of sensors is $10$ (see Fig.~\ref{exp_co_ap} in Section \ref{sec:decoded_packcet3} for the details).

In Co-AP, AP2 can help AP1 by forwarding an update packet only when the packet is successfully decoded. That is, AP2 does not help if it fails to decode the update packet either. However, failure to decode the packet does not mean that there is no information about the packet. Intuitively, AP2 can forward the received raw complex signals to AP1. Then AP1 tries to decode the packet again jointly using the signals received at both APs, i.e., by the maximum ratio combining (MRC) technique \cite{A. Goldsmith}. While forwarding raw signal samples improves the joint decoding performance, it requires AP2 to forward a large amount of data. This may affect the regular traffic over the backbone network, e.g., the backbone network infrastructure among lamp posts in smart cities is shared by other networks. Moreover, forwarding a large amount of data increases the backbone delay, i.e., the time for AP1 to receive the raw samples from AP2. Thus, the average AoI may still be high, even if the update packet is eventually decoded at AP1.


Modern channel decoding methods often utilize soft information on the coded bits of a packet that give the log-likelihood ratio (LLR) of the probabilities of the bits being zero and one, referred to as soft bits \cite{ShuCoding}. Built upon Co-AP, we further investigate an improved Co-AP system, referred to as the Soft-Co-AP system. Instead of forwarding raw signals, Soft-Co-AP forwards soft bits of the coded packet from AP2 to AP1, when AP2 fails to decode the packet. With the help of soft bits, even if both packet copies fail to be decoded at the APs, the packet can still have a chance to be decoded after joint decoding at AP1. For fast decoding, the LLRs of the coded packet at AP2 generally need to be discretized and quantized from real numbers to integers before forwarding to the channel decoder in many practical systems. For example, a soft bit can be represented by eight quantization bits in our adopted SDR implementation \cite{Spiral Project}, i.e., a soft bit is an integer ranging from $0$ to $255$, with $0$ ($255$) being most likely to be a coded bit of $0$ ($1$). This further reduces the amount of data forwarded over the backbone network. 

Intuitively, the more quantization bits used to represent a soft bit, the better the joint decoding performance at AP1, i.e., using more quantization bits loses less information about the original real-value LLR \cite{ShuCoding}. However, more quantization bits used lead to a higher backbone delay, affecting the average AoI. In contrast, using fewer quantization bits leads to a lower backbone delay, but degrades the joint decoding performance. Hence, the impact of the number of quantization bits on the average AoI requires an in-depth investigation in a backbone-assisted network considered in this paper. 

Finally, we conduct experiments with real Ethernet backbone delay to compare the average AoI between Co-AP and Soft-Co-AP. Furthermore, we study the average AoI of Soft-Co-AP under different numbers of quantization bits and network configurations. Our SDR experimental results show that when the number of sensors is large (say, $30$ sensors), Soft-Co-AP further reduces the average AoI of Co-AP by $12$\%. In addition,  we observe from our experiments that four quantization bits per soft bit are usually enough to achieve the best average AoI performance of Soft-Co-AP, owing to the tradeoff between the joint decoding performance and the backbone delay. Overall, Soft-Co-AP is a viable solution for improving information freshness with massive IoT sensors.

To sum up, there are three major contributions in this paper:
\begin{enumerate}
    \item We are the first to study backbone-assisted cooperative APs to improve information freshness. Our SDR experiments show that a simple Co-AP system wherein only decoded packets are forwarded through the backbone significantly reduces the average AoI of the network by $50$\%, compared with a conventional single-AP system.
    \item We further investigate an improved Co-AP system, called Soft-Co-AP. Soft-Co-AP can forward the soft bits of update packets that are failed to be decoded. While more quantization bits per soft bit improve the successful decoding probability of packets, a more considerable backbone delay is induced. Since both successful decoding probability and backbone delay affect the average AoI performance, we experimentally explore the impact of the number of quantization bits per soft bit on the average AoI in Soft-Co-AP.
    \item We demonstrate the practical feasibility of status update systems with backbone-assisted cooperative APs. SDR experimental results indicate that the average AoI performance of Soft-Co-AP outperforms Co-AP and single-AP significantly, especially when the number of sensors is large. Interestingly, the number of quantization bits per soft bit in Soft-Co-AP is usually neither too large nor too small to achieve a good average AoI performance.  
\end{enumerate}

The rest of this paper is organized as follows.
Section \ref{sec:relatedwork} details related work on AoI and compares Co-AP/Soft-Co-AP with relevant techniques in the existing literature. Section  \ref{sec:decoded_packcet} describes the system model of Co-AP and previews experimental results, showing that Co-AP reduces average AoI compared with the single-AP system. In Section \ref{sec:soft_bits}, we present Soft-Co-AP and explain the joint decoding mechanism with the help of soft bits. Practical experimental evaluations of the average AoI in different systems are discussed and compared in Section \ref{sec:exp}. Finally, conclusions are drawn in Section \ref{sec:conl}. Table \ref{tab:notation} summarizes the key acronyms and notations of this paper. 

\section{Related Work} \label{sec:relatedwork}
To characterize the freshness of the received information at the destination, AoI was introduced in vehicular networks \cite{S. Kaul}. Later, AoI was investigated in various wireless communication scenarios, such as industrial IoT \cite{X. Wang,H. Pan2}, health monitoring \cite{L. Guo,Z. Ling}, and satellite networks\cite{Y. Li}. At the beginning of the study of AoI, the queuing theory was widely used to investigate the age performance under different abstract queuing models \cite{A. Kosta,Y. Inoue,A. M. Bedewy}. In addition, different user scheduling algorithms were proposed, aiming to minimize different age metrics, such as average AoI \cite{A. Kosta} and peak AoI \cite{J. P. Champati}. These works were usually theoretical in nature and based on an ideal assumption that the status update is sent through a perfect channel without packet errors.  

Recent studies on AoI focus more on the lower layer of the communication stack, i.e., PHY and MAC layers, taking unreliable wireless channels into account. To mitigate the wireless impairments that cause packet errors or loss, different PHY-layer techniques are proposed to reduce the average AoI. For example, \cite{M. Xie, E. T. Ceran} focused on packet retransmission in which automatic repeat request (ARQ) or  hybrid ARQ (HARQ) is used to improve the packet reception probability. Packet management techniques on the old packets and the newly generated packets at the transmitter were carefully designed to improve information freshness \cite{H. Pan3, Costa. M, M. Xie2}. Besides, age-aware channel coding techniques were proposed in \cite{X. Chen, H. Pan2}, indicating that optimizing the average AoI usually leads to different designs of coding redundancy and coding strategies. 

Conventional multiple-input-multiple-output (MIMO) systems increase spatial diversity by equipping with more than one antenna at the receiver \cite{A. Goldsmith}. MIMO systems were investigated in \cite{S. Feng,B. Yu} to improve the AoI performance. The backbone-assisted cooperative-AP system considered in this paper use multiple receivers to exploit spatial diversity. Hence, it shares the same idea as MIMO systems that additional degrees of freedom increase the packet reception probability. However, traditional MIMO systems, where multiple antennas are mounted on the same receiver, are different from our cooperative-AP scenario, where APs are interconnected with a backbone network. Specifically, as presented in this work, the backbone delay therein does affect the AoI performance, which does not occur in conventional MIMO systems.

\begin{table}[]
\caption{List of Key Acronyms and Notations}
	\centering
\begin{adjustbox}{width=\columnwidth,center}
	{\color{black}{\begin{tabular}{r l}
	\hline
			AoI : & \ age of information    \\ 
			AP : & \ access point    \\ 
			AWGN : & \ additive white
			Gaussian noise    \\ 
			BPSK : & \ binary phase shift keying    \\ 
			CP : & \ cyclic prefix   \\ 
			FFT : & \ fast Fourier transform   \\ 
			IFFT : & \ inverse fast Fourier transform   \\ 
			LLR : & \ log-likelihood ratio    \\
			MAC : & \ media access control    \\
			MIMO : & \ multiple-input-multiple-output \\
			MRC :& \ maximum ratio combining     \\ 
			OFDM : & \  orthogonal frequency-division multiplexing    \\
		    SDR : & \ software-defined radio    \\
		    SNR : & \ signal-to-noise ratio    \\
		    TDMA : & \  time division multiple access \\ 
		    UHD : & \  USRP hardware driver \\
		    VA : & \  Viterbi decoding algorithm \\
		    USRP : & \  universal software radio peripheral \\
		    \hline 
			$N$  : & \ the number of sensors    \\
            ${T}$       :&\ time slot duration \\  
            $C_j^i$      :& \ the $j$-th update packet generated by sensor $i$  \\
            $t_j^i$      :& \ generation time of the $j$-th update packet of sensor $i$\\
			${\Delta _i}(t)$      : & \  instantaneous AoI of sensor $i$ at time $t$   \\  
			${U_i}(t)$   :    & \ generation time of the latest update received from sensor $i$ at time $t$\\  
			${\bar \Delta _i}$       : &\ average AoI of sensor $i$ \\ 
			$Z_i(w)$         :&\ time required for the $w$-th update since the $(w$-$1)$-th update \\  
			$t_{delay}$      :&\ backbone delay \\  
			$V^i$    :&\ binary codeword of packet $C^i$ after channel coding \\  
			$v_i[k]$    :&\ the $k$-th coded bit of $V^i$\\
			$X$ :&\ BPSK modulated symbols  of $V^i$      \\  
			$x[k]$ :&\  the $k$-th modulated BPSK symbol of $X$ \\ 
			$Y_r$       :&\ received symbols at AP $r$, $r\in\{1,2\}$ \\
            $y_r[k]$   :&\ the $k$-th received symbol at AP $r$, $r\in\{1,2\}$ \\
            $h_r[k]$   :&\ channel gain of the $k$-th symbol at AP $r$, $r\in\{1,2\}$\\
            $n[k]$   :&\ AWGN of the $k$-th symbol at AP $r$, $r\in\{1,2\}$\\
            $\widetilde x_r[k]$ :&\ LLR of the coded bit $v_i[k]$ at AP $r$, $r\in\{1,2\}$\\
			$\widetilde x_{r,q}[k]$  :&\ $8$-bit quantized soft bit of $v_i[k]$ at AP $r$, $r\in\{1,2\}$\\
			$\widetilde x^{co}[k]$       :&\ LLR of $v_i[k]$ when jointly considering two APs by MRC    \\ 
            $\widetilde x_{r,q,m}[k]$       :&\ $m$-bit quantized soft bit at AP $r$, $r\in\{1,2\}$    \\
            $\widetilde x_q^{co}[k]$       :&\  combined quantized soft bit of $v_i[k]$ in Soft-Co-AP\\
		    
 \hline
 \hline
	\end{tabular}}}
	\end{adjustbox}
	\label{tab:notation}
\end{table}

Our backbone-assisted cooperative-AP systems, namely Co-AP and Soft-Co-AP, are also similar to the distributed MIMO systems studied in \cite{K. Tan, H. Balan}. Coordinated multipoint (CoMP) \cite{Irmer, D. Lee} and cell-free massive MIMO \cite{Ammar} are examples of distributed MIMO systems where spatially separated transmitters form a virtual MIMO system for multiple access. Applying previous works on CoMP or cell-free massive MIMO to status update systems requires cooperative APs to exchange raw signals for joint decoding. However, this raw signal exchange is typically accomplished by backhaul networks connected with dedicated high-speed fibers. When connected via lower-cost Ethernet, previous research \cite{W. Zhou, H. Pan4} pointed out that exchanging raw signal samples over the Ethernet backbone could lead to unaffordable traffic. More importantly, the high backbone delay induced is detrimental to low-AoI networks. Besides, \cite{G. Woo, M. Gowda} also exploited the soft information from multiple APs to improve decoding performance as Soft-Co-AP does. However, all the above works on cooperative receivers focused on boosting system throughput rather than reducing AoI. By contrast, this paper examines the design of soft bits (i.e., the number of quantization bits representing each coded bit shared over the backbone) on the average AoI. We experimentally show that good average AoI performance is usually achieved when the number of quantization bits per soft bit is neither too large nor too small. Such results also indicate that exchanging raw samples as in CoMP or cell-free MIMO does not reduce the average AoI due to the high backbone delay induced when exchanging a large volume of raw signal samples.

\section{Cooperative Access Points I: Forwarding Decoded Packets} \label{sec:decoded_packcet}
We now detail the status update system architecture with backbone-assisted cooperative APs. Specifically, this section focuses on a cooperative scheme in which only the decoded packets are shared by an Ethernet backbone connecting the cooperative APs. We show that sharing the decoded packets through the Ethernet backbone improves information freshness. Section \ref{sec:soft_bits} further studies forwarding the soft bits of coded packets when the packets are not successfully decoded at the APs. In both Sections \ref{sec:decoded_packcet} and \ref{sec:soft_bits}, we focus on two cooperative APs for easy illustration. Generalization from two APs to multiple APs is straightforward. 

\subsection{System Model}\label{sec:decoded_packcet1}
Considering a dense AP scenario, we focus on a status update system where two neighboring APs, namely AP1 and AP2, serve multiple sensors located in the overlapping coverage area of the two APs, as shown in Fig.~\ref{system_model}(b). A total of $N$ sensors want to send update packets to AP1. Let us assume that AP1 is the primary AP and AP2 is the secondary AP. In practice, each AP could be primary or secondary to different sensors, which can be directly generalized from our current example. Both AP1 and AP2 receive the wireless signals sent by the sensors and try to decode their update packets. An Ethernet backbone connection is established between AP1 and AP2 so that information can be forwarded from the secondary AP (AP2) to the primary AP (AP1). In this section, the information forwarded over the backbone is restricted to decoded packets only. We refer to the above system architecture as the backbone-assisted COoperative-AP (Co-AP) system.

Throughout this paper, we assume that the $N$ sensors send their update packets in a TDMA manner, where time is divided into a series of TDMA rounds. A TDMA round consists of $N$ time slots with the same duration $T$. Different sensors send their packets in a round-robin manner, and each sensor occupies a time slot. Note that Co-AP applies to other channel access schemes such as carrier-sensing multiple access (CSMA) used in the IEEE 802.11 standards\cite{IEEE802.11ac} as well. 

We further assume a generate-at-will packet generation model in which the update packet about the observed phenomena can be generated  when the sensor has the transmission opportunity \cite{M. A. Abd-Elmagid}. Following the generate-at-will model, in each TDMA round, a sensor generates and sends a new update packet just at the beginning of its allocated time slot. This ensures that the sampled information is as fresh as possible, i.e., a sensor reading is obtained just before its transmission opportunity. More specifically, with respect to Fig.~\ref{aoi_examples}, sensor \emph{i} generates and sends packet $C_{j}^i$ at $t_j^i$ in TDMA round \emph{j} and completes the transmission at $t_j^{i+1}$, where $t_j^{i+1}=t_j^i+T$. Both AP1 and AP2 try to decode $C_j^i$ at $t_j^{i+1}$. If AP2 decodes $C_j^i$ successfully, it forwards $C_j^i$ to AP1 through the Ethernet backbone in Co-AP. The evolution of AoI is then described in detail below.

\subsection{Age of Information (AoI)} \label{sec:decoded_packcet2}
In the status update system shown in Fig.~\ref{system_model}(b), the primary AP, AP1, wants to receive the update packets from sensors as fresh as possible. This paper adopts AoI to quantify the information freshness. At any time $t$, the instantaneous AoI of sensor $i$, $i \in \{ 1,2,...,N\}$, measured at the primary AP (i.e., AP1), is defined by
\begin{align}
{\Delta _i}(t) = t - {U_i}(t),
\end{align}
where \emph{${U_i}(t)$} is the generation time of the last successfully received update packet of sensor $i$ at AP1. The smaller the instantaneous AoI ${\Delta _i}(t)$, the more recent the information from sensor $i$. With the instantaneous AoI $\Delta_i (t)$, we can compute the \emph{average AoI}, which measures the time average of the instantaneous AoI. The average AoI of sensor $i$, ${\overline \Delta  _i} $, is defined by
\begin{align}
{\overline \Delta  _i} = \mathop {\lim }\limits_{T \to \infty } \frac{1}{T}\int_0^T {{\Delta _i}(t)} dt.
\end{align}

\begin{figure}[t]
\centerline{\includegraphics[trim=15 0 0 0,width=0.5\textwidth]{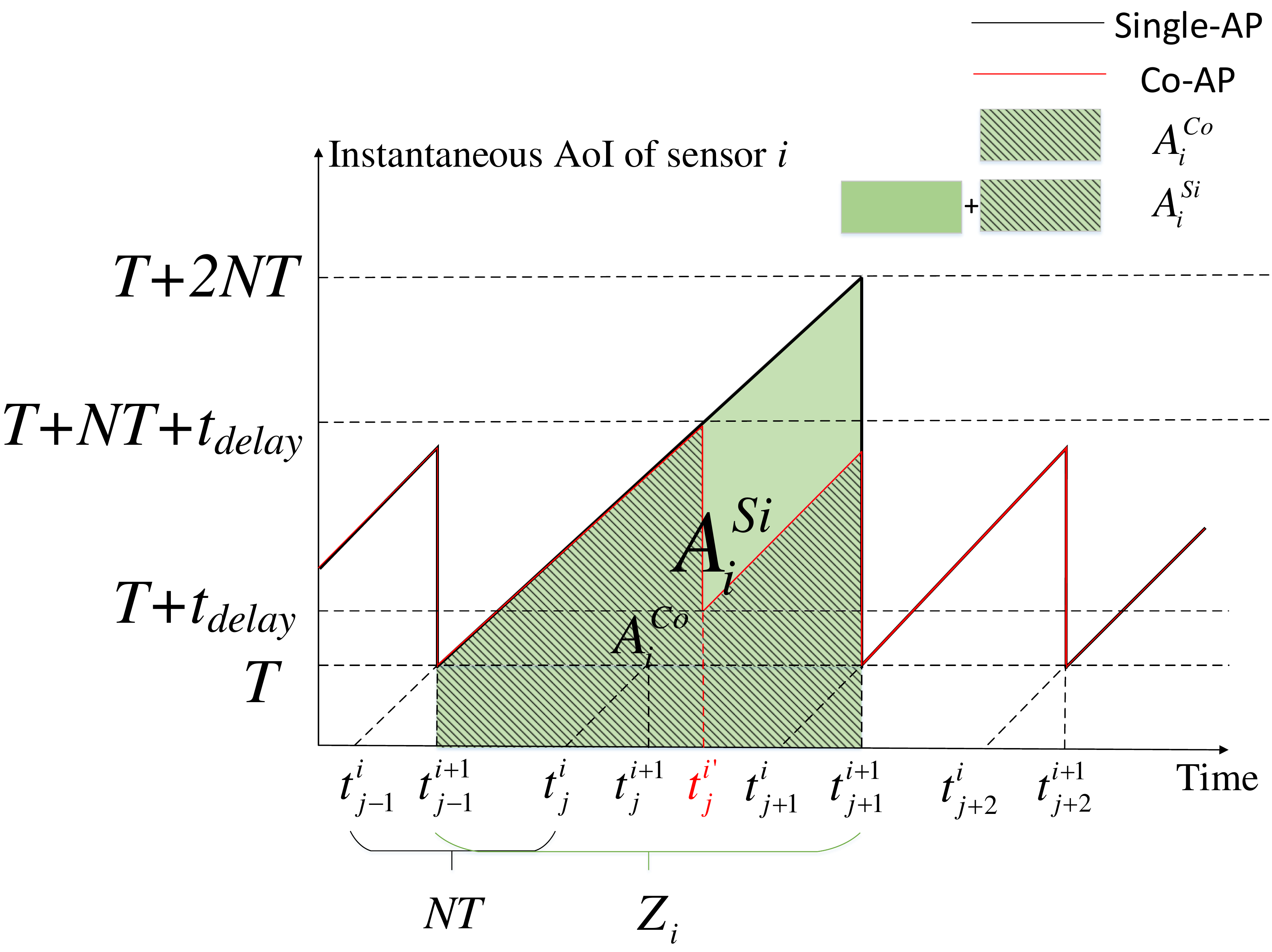}}
\caption{An example of instantaneous AoI of sensor \emph{i} when using single-AP or Co-AP respectively.}
\label{aoi_examples}
\end{figure}

We now focus on an example shown in Fig.~\ref{aoi_examples} to see how cooperative APs help reduce the average AoI, compared with conventional systems with non-cooperative APs. If APs do not cooperate, AP2 does not forward the decoded packets to AP1, i.e., the system is in fact a \emph{single-AP} system. Let us first consider the instantaneous AoI of sensor $i$ in single-AP, as depicted by the black curve in Fig.~\ref{aoi_examples}. In Fig.~\ref{aoi_examples}, sensor $i$ sends update packets $C_{j-1}^i$, $C_j^i$, and $C_{j+1}^i$ at times $t_{j-1}^{i}$, $t_{j}^{i}$, and $t_{j+1}^{i}$, respectively. Only packets $C_{j-1}^i$ and $C_{j+1}^i$ are received successfully by AP1. Hence, the instantaneous AoI is reduced to $T$ at times $t_{j-1}^{i+1}$ and $t_{j+1}^{i+1}$. When AP1 fails to decode packet $C_j^i$, the instantaneous AoI continues to increase. In other words, between the two consecutive updates at times $t_{j-1}^{i+1}$ and $t_{j+1}^{i+1}$, the instantaneous AoI has to increase linearly, resulting in a trapezoidal area $A^{Si}_i$ for user $i$ under the instantaneous AoI curve, as shown in Fig.~\ref{aoi_examples}. We refer to this area as the AoI area. 

The average AoI is calculated by accumulating a series of AoI areas, divided by the total time \cite{A. Kosta}, i.e.,
\begin{align}
{\overline \Delta  _i} = \mathop {\lim }\limits_{T \to \infty } \frac{1}{T}\int_0^T {{\Delta _i}(t)} dt = \mathop {\lim }\limits_{W \to \infty } \frac{{\sum\nolimits_{w = 1}^W {{A_i^{Si}(w)}} }}{{\sum\nolimits_{w = 1}^W {{Z_i(w)}} }},
\label{equ:aoi}
\end{align}
where $A_i^{Si}(w)$ is the $w$-th AoI area and $Z_i(w)$ is the time required for the $w$-th update since the $(w-1)$-th update. It is obvious that during the same time period, a smaller AoI area results in a smaller average AoI. 
 
Now we consider a  Co-AP system. The instantaneous AoI of sensor $i$ is depicted by the red curve in Fig.~\ref{aoi_examples}. Suppose that AP1 fails to decode packet $C_j^i$ at $t_{j}^{i+1}=t_{j}^{i}+T$, but AP2 decodes $C_j^i$ successfully. Then AP2 forwards the decoded $C_j^i$ to AP1 through the Ethernet backbone. In Fig.~\ref{aoi_examples}, we assume that AP1 receives $C_j^i$ forwarded by AP2 after a backbone delay $t_{delay}$, i.e., AP1 receives $C_j^i$ at ${t_j^{i}}'=t_{j}^i+T+t_{delay}$. Hence, the instantaneous AoI of sensor $i$ can be reduced to $T+t_{delay}$ at ${t_j^{i}}'$. Compared with single-AP, thanks to the update packet forwarded by AP2, the instantaneous AoI in Co-AP can be reduced before the next update opportunity at $t_{j+1}^{i+1}$. This leads to a much smaller AoI area under the instantaneous AoI curve. Based on (\ref{equ:aoi}), a smaller AoI area leads to a smaller average AoI given the same time duration. 

From the above example, we observe that in Co-AP, the decoded packets forwarded by AP2 can reduce the average AoI as long as the backbone delay $t_{delay}$ is small enough. For example, in Fig.~\ref{aoi_examples}, AP1 should receive $C_j^i$ from AP2 before $t_{j+1}^{i+1}$ (this is the time of the next update packet decoded by AP1 itself) so as to reduce the time interval between two consecutive updates. In practical Ethernet backbone environments, the backbone delay is random and depends on the backbone traffic. The merits of backbone-assisted Co-AP systems should be validated in real wireless systems, as will be shown later in this paper.

\begin{figure}[t]
\centerline{\includegraphics[trim=0 0 0 0,width=0.5\textwidth]{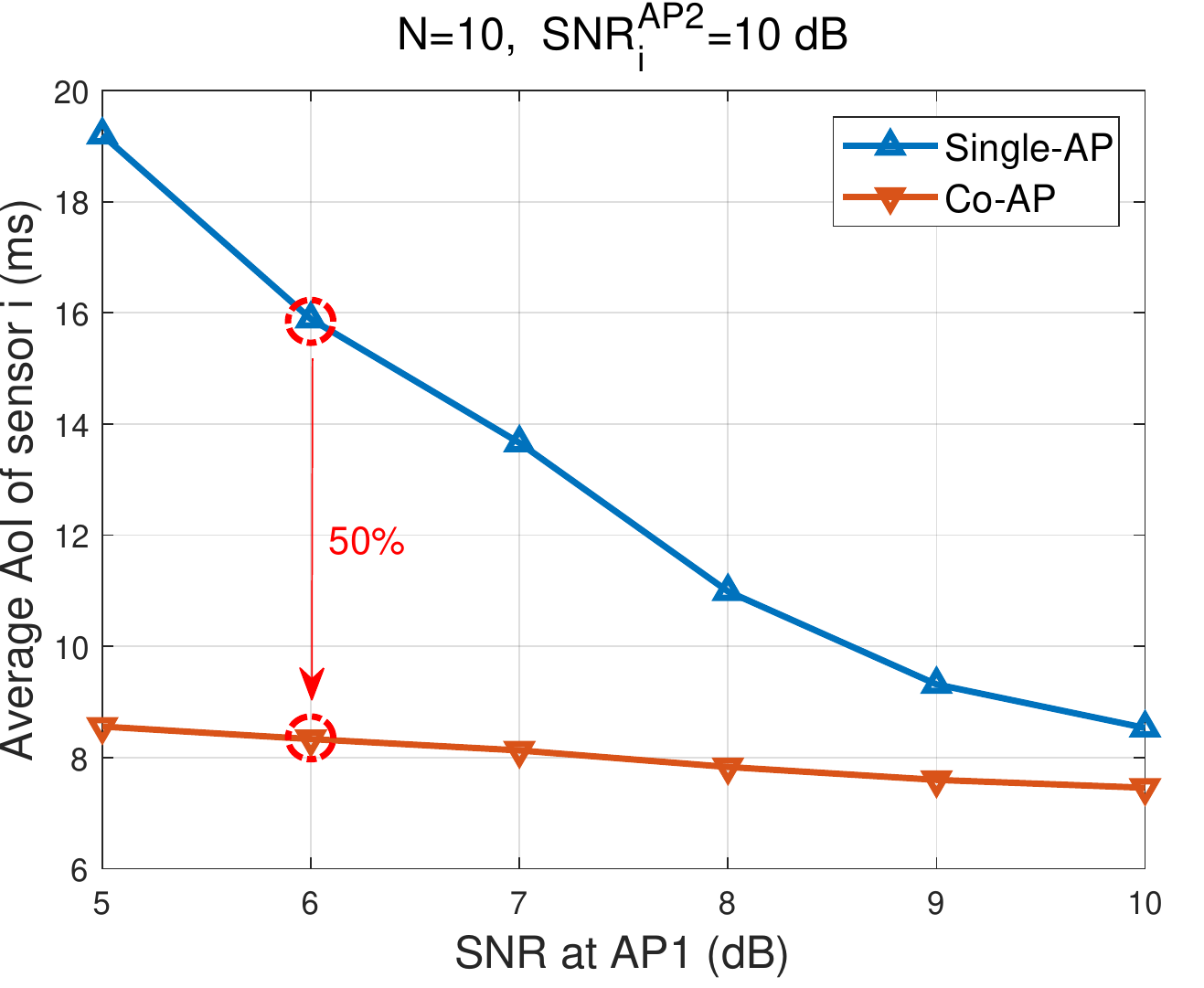}}
\caption{Experimental average AoI of sensor $i$ in single-AP and Co-AP systems when the number $N$ of sensors is $10$ and $SNR_i^{AP2}=10$dB ($SNR_i^{AP2}$ means sensor $i$'s SNR at AP2). }  
\label{exp_co_ap}
\end{figure}

\subsection{Co-AP Reduces Average AoI } \label{sec:decoded_packcet3}
Let us preview some experimental results on SDR \cite{G. FSF} and compare the average AoI between single-AP and Co-AP. In this experiment, we assume that the number of sensors is $10$, and they have the same received SNR at the APs. The detailed experimental setup and results can be found in Section \ref{sec:exp}. We deploy the Universal Software Radio Peripheral (USRP) devices \cite{Ettus} in an indoor environment and conduct experiments on our prototype. 

We conduct trace-driven simulations. Specifically, as will be detailed in Section \ref{sec:exp1}, we first obtain the physical (PHY)-layer decoding outcomes using SDR, i.e., collecting the decoding results at the two APs, and then use the traces to create events in AoI simulations. In this experiment, we vary the SNR of all sensors at AP1 ($SNR^{AP1}$) from $5$dB to $10$dB and fix the SNR at AP2 ($SNR^{AP2}$) to $10$dB. As such, sensors have better channel conditions at AP2 than at AP1. Also, note that the backbone delay statistics are collected from the real Ethernet backbone (see Section \ref{sec:exp1} for the backbone delay measurement setup). 

Fig.~\ref{exp_co_ap} plots the average AoI of one sensor (say, sensor $i$) versus $SNR^{AP1}$. We see in Fig.~\ref{exp_co_ap} that Co-AP has a significantly lower average AoI than single-AP does, especially when the received SNR difference between the two APs is large. For example, when $SNR^{AP1}=6$dB and $SNR^{AP2}=10$dB, Co-AP reduces the average AoI of sensor $i$ by around $50$\% compared with single-AP. Thanks to the decoded packets forwarded by AP2, the instantaneous AoI of sensor $i$ can be dropped before the update opportunity in the next TDMA round, thus achieving higher information freshness. 

Notice that in Co-AP, AP2 forwards an update packet to AP1 only when the packet is decoded successfully. When AP2 cannot decode the packet, intuitively, it can forward the raw complex signal samples to AP1 so that AP1 can try to decode the packet again by jointly using the signals received at both APs. However, forwarding raw signal samples leads to a large amount of backbone traffic and a high backbone delay \cite{W. Zhou, H. Pan4}, thus affecting the AoI performance. To avoid high backbone delay and further improve the average AoI performance, a promising solution is to forward the soft bits of the coded packet for joint decoding at the primary AP. Section \ref{sec:soft_bits} below considers an improved Co-AP system where the secondary AP can forward soft bits to the primary AP to help with status updates.

\begin{figure}[t]
\centerline{\includegraphics[trim=0 0 0 0,width=0.5\textwidth]{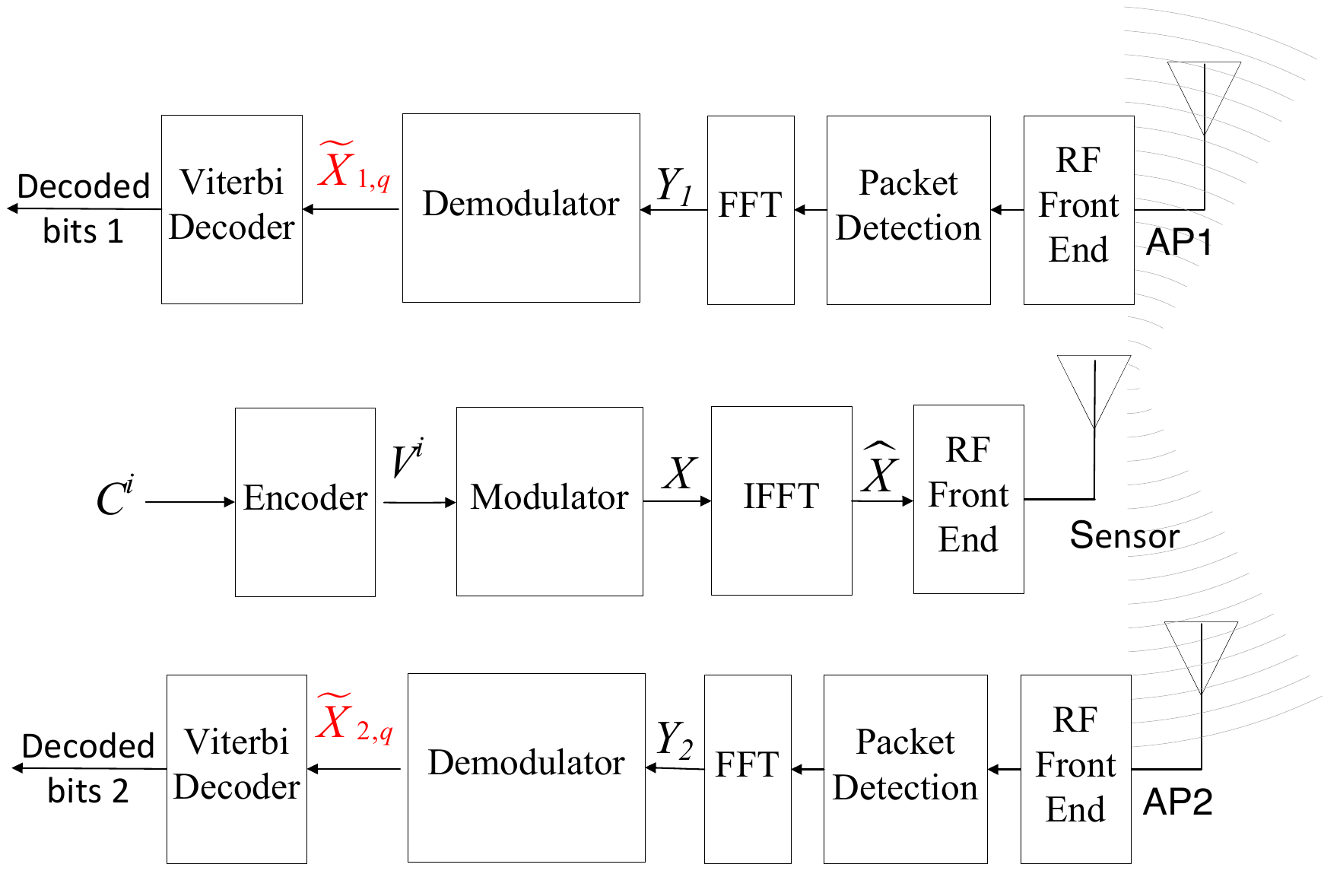}}
\caption{The structure of sensor $i$'s transmitter and  both APs' receivers operating during one time slot in a Co-AP system.}  
\label{soft_co_ap_structure}
\end{figure}

\section{Cooperative Access Points II: Forwarding Soft Bits }\label{sec:soft_bits}
In this section, we present an improved Co-AP system, referred to as the Soft-Co-AP system. Comparing with Co-AP, Soft-Co-AP adds a mechanism that the secondary AP can forward the soft bits of a coded packet to the primary AP through the Ethernet backbone. In other words, when AP2 fails to decode an update packet, AP2 forwards the soft bits of that packet to AP1. AP1 tries to recover the update packet it previously failed to decode using its own received samples and the soft bits forwarded by AP2. Section \ref{sec:soft_bits1} describes the overall system architecture and explains the soft bits forwarded by AP2. After that, Section \ref{sec:soft_bits2} presents the joint decoding mechanism with the help of soft bits. In particular, we explain the impact of soft bits on the average AoI. 

\subsection{Overall Soft-Co-AP System Architecture}\label{sec:soft_bits1}
As illustrated in Fig.~\ref{soft_co_ap_structure}, to send an update packet, sensor $i$ first encodes a source update packet $C^i$ (for simplicity, we drop the TDMA round index $j$) to the binary codeword $V^i=(v_i[1],...,v_i[k],...)$, where $v_i[k]\in \{0, 1\}$ is the $k$-th coded bit of $C^i$. In this paper, we assume the use of convolution codes defined in the IEEE 802.11 standards \cite{IEEE802.11ac} as the PHY-layer channel code to improve transmission reliability. The codeword $V^i$ is then BPSK-modulated into the BPSK symbols ${ X=( x[1],..., x[k],...)}$, where $ x[k] = 1-2v_i[k]$. Extensions to higher-order modulations beyond BPSK are straightforward. As in 802.11, our system uses orthogonal frequency division multiplexing (OFDM) at the PHY layer. Specifically, inverse fast Fourier transform (IFFT) is used so that BPSK symbols ${\widehat X}$ are modulated into ${\widehat X=(\widehat x[1],...,\widehat x[k],...)}$ at the transmitter, which are sent to both APs. In addition, a cyclic prefix (CP) is inserted for each OFDM symbol to deal with multi-path fading \cite{A. Goldsmith}. 

Both AP1 and AP2 receive the signals sent by sensor $i$, i.e., ${ Y_1=( y_1[1],..., y_1[k],...)}$ and ${ Y_2=( y_2[1],..., y_2[k],...)}$ are frequency-domain signals after FFT at the receiver, respectively, where $y_r[k]$ is the $k$-th symbol received at AP $r\in\{1,2\}$. Assuming that the multi-path effect can be dealt with by the CPs of OFDM symbols, $y_r[k]$ is given by
\begin{equation}
{y_r[k]} = {h_{r}}[k]x[k]+n[k],
\end{equation}
where $x[k]$ is the $k$-th transmitted BPSK symbol, $n[k]$ is the additive white Gaussian noise (AWGN) with variance $\sigma ^2$, and $h_{r}[k]$ is the channel gain of symbol $x[k]$ at AP $r$.

Both APs adopt the soft-input Viterbi decoding algorithm (VA) to decode the source packet $C^i$. Specifically, soft bits are computed from the received symbols $Y_r$ at AP $r$, which are then fed to the Viterbi decoder to decode $C^i$. The idea of the soft-input VA is to provide a confidence metric $\widetilde x_r[k]$ of AP $r$ for each coded bit $v_i[k]$ to the Viterbi shortest-path algorithm. The confidence metric $\widetilde x_r[k]$ is referred to as the soft bit of $v_i[k]$ and is computed by the log-likelihood ratio (LLR), i.e.,
\begin{align}
\widetilde x_r[k]&=\log \frac{{{P_0}[k]}}{{{P_1}[k]}} \notag \\
&= \frac{{{1}}}{2\sigma ^2} ( - {\left| {{y_r}[k] - {h_r}[k]} \right|^2} + {\left| {{y_r}[k] + {h_r}[k]} \right|^2}) \notag \\
&= \frac{{{2}}}{\sigma ^2}~{y_r}[k] \cdot {h_r}[k] \\
&\propto {y_r}[k] \cdot {h_r}[k],
\end{align}
where 
\begin{align}
&P_0[k]=\Pr(x[k]=1|{y_r[k]})
=
\frac{1}{{\sqrt {2\pi {\sigma ^2}} }}{e^{\frac{{ - |{y_r}[k] - {h_r}[k]{|^2}}}{{2{\sigma ^2}}}}},\\
&P_1[k]=\Pr(x[k]=-1|{y_r[k]})=
\frac{1}{{\sqrt {2\pi {\sigma ^2}} }}{e^{\frac{{ - |{y_r}[k] + {h_r}[k]{|^2}}}{{2{\sigma ^2}}}}}.
\end{align}
The dot ``~$\cdot$~'' in ($5$) represents the dot product. We denote the two complex numbers ${y_r}[k]$ and ${h_r}[k]$ as vectors with two elements, one of which is the real part of the corresponding complex number and one of which is the imaginary part. From ($5$) to ($6$), we remove the constant term $\frac{{{2}}}{\sigma ^2}$ since the constant term does not affect the shortest path found by VA.

For easy implementation and fast decoding, the soft-input Viterbi decoder adopted in the many practical systems (including the SDR platform used in this paper) accepts confidence metrics represented by integers from $0$ to $255$. Since $\widetilde x_r[k]$ is a real value, it needs to be quantized before feeding to the Viterbi decoder, i.e., $\widetilde x_r[k]$ needs to be quantized into $\widetilde x_{r,q}[k]$. For simplicity, let us consider a noiseless case to explain the quantization procedure. According to ($6$), we have
\begin{align}
\widetilde x_{r}[k]  \propto {y_r}[k] \cdot {h_r}[k] =x[k]\times |h_r[k]|^2,
\end{align}
where $x[k] \in \{1,-1\}$. That is, $\widetilde x_{r}[k]\propto \pm |h_r[k]|^2$. Over all $k$, we focus on the constellation points with the largest magnitude, which is given by
\begin{align}
|{h_{r, \max }}{|^2} = \mathop {\max }\limits_k |{h_r}[k]{|^2}.
\end{align}
As in \cite{L. Lu}, we quantize $\widetilde x_{r}[k]$ to $\widetilde x_{r,q}[k]$ by 
\begin{align}
\widetilde x_{r,q}[k] = \left(\frac{{ {\widetilde x_r[k]}}}{{|{h_{r,\max }}{|^2}}}\alpha  + \beta \right) \times 255.
\end{align}
That is, we first map $\pm |h_r[k]|^2$ to two points falling in the interval $(-1,1)$. The parameters $\alpha$ and $\beta$ are used to map the interval $(-1,1)$ to a new interval roughly from $0$ to $1$ so that $\widetilde x_{r,q}[k]$ can fall between $0$ and $255$ after multiplying 255. In practice, $\alpha  \in [0.2,0.5]$ and $\beta=0.5$. As a result, if $\widetilde x_{r,q}[k]$ is closer to 255 (or 0), $v_i[k]$ is more likely to be a coded bit of 1 (or a coded bit of 0) \cite{L. Lu}. 

The quantized soft bits ${ \widetilde X_{1,q}=( \widetilde x_{1,q}[1],..., \widetilde x_{1,q}[k],...)}$ and ${ \widetilde X_{2,q}=( \widetilde x_{2,q}[1],..., \widetilde x_{2,q}[k],...)}$ are fed into the Viterbi decoder at AP1 and AP2, respectively, to decode $C^i$. If AP2 fails to decode $C^i$, soft bits are forwarded to AP1 by the Ethernet backbone and AP1 tries to decode $C^i$ again. The joint decoding mechanism with the help of soft bits will be discussed in the next subsection. 

\subsection{Joint Decoding with the Forwarded Soft Bits}\label{sec:soft_bits2}

Since each soft bit $\widetilde x_{r,q}[k]$ is an integer from $0$ to $255(=2^8-1)$, it requires $8$ quantization bits to represent each $\widetilde x_{r,q}[k]$, $r\in\{1,2\}$. As measured by our backbone delay experiments (see Section \ref{sec:exp1} for the details), when forwarding the soft bits to AP1, using $8$ quantization bits for each $\widetilde x_{r,q}[k]$ leads to a high average backbone delay of $20ms$. In general, forwarding more quantization bits per soft bit through the Ethernet backbone induces higher backbone delay, thereby affecting the AoI performance. However, forwarding fewer quantization bits per soft bit loses information (i.e., a coarse quantization) and degrades the joint decoding performance at the primary AP. Next, we investigate forwarding fewer quantization bits per soft bit to AP1 and study how to combine the information from both APs to jointly decode update packets.

Suppose that backbone delay is not a concern. AP2 should forward the received signal ${y_2[k]}$ to AP1 to achieve the optimal decoding performance. Specifically, AP1 outputs the confidence metric $\widetilde {x}^{\rm{co}}[k]$ by jointly considering ${y_1[k]}$ and ${y_2[k]}$, i.e., $\widetilde {x}^{\rm{co}}[k]$ is given by  
\begin{align}
\widetilde {x}^{\rm{co}}[k]&=\log \frac{{{P_0^{\rm{co}}}[k]}}{{{P_1^{\rm{co}}}[k]}} = \log{{P_0^{\rm{co}}}[k]} - \log{{P_1^{\rm{co}}}[k]}, \notag \\
&\propto \widetilde x_1[k] + \widetilde x_2[k],
\end{align}
where 
\begin{align}
P_0^{\rm{co}}[k]&=\Pr(x[k]=1|{y_1[k],y_2[k]}) \notag \\
&\propto {e^{\frac{{ - |{y_1}[k] - {h_1}[k]{|^2}}}{{2{\sigma ^2}}}}}{e^{\frac{{ - |{y_2}[k] - {h_2}[k]{|^2}}}{{2{\sigma ^2}}}}},
\end{align}
\begin{align}
P_1^{\rm{co}}[k]&=\Pr(x[k]=-1|{y_1[k],y_2[k]})\notag \\
&\propto {e^{\frac{{ - |{y_1}[k] + {h_1}[k]{|^2}}}{{2{\sigma ^2}}}}}{e^{\frac{{ - |{y_2}[k] + {h_2}[k]{|^2}}}{{2{\sigma ^2}}}}}.
\end{align}
Here, we assume the two APs have the same noise power. From ($12$), we see that the optimal way to compute the LLR of bit $v_i[k]$ is to add the soft bits $\widetilde x_1[k]$ and $\widetilde x_2[k]$ directly. Given that AP1 receives a quantized version of $\widetilde x_2[k]$ from AP2, we next present how to compute the LLR of bit $v_i[k]$ using the quantized version of $\widetilde x_2[k]$. 

Let $m$ denote the number of quantization bits to represent $\widetilde x_2[k]$. Suppose that a $m$-bit quantized version of $\widetilde x_2[k]$ is represented by $\widetilde x_{2,q,m}[k]$, $m\in\{1,2,...,8\}$. Note that $\widetilde x_{2,q}[k]$ in ($11$) is a special case $\widetilde x_{2,q,m}[k]$ with $m=8$. We consider uniform quantization to quantize from $\widetilde x_{2,q}[k]$ to $\widetilde x_{2,q,m}[k]$. With $m\in\{1,2,...,8\}$, the number of quantization levels is $2^m$, i.e., the $2^m$ levels are 
\begin{align}
\left\{0, \lfloor 1\times \frac{255}{2^m-1}\rfloor, \lfloor 2\times\frac{255}{2^m-1}\rfloor ,...,{(2^m-1)}\times\frac{255}{2^m-1}\right\},
\end{align}
where $\lfloor \cdot \rfloor$ is the round-up operator to convert the values of quantization levels to integers. Hence, $\widetilde x_{2,q,m}[k]$ can be obtained by quantizing $\widetilde x_{2,q}[k]$ to the nearest quantization level. For example, when $m=1$, there are two quantization levels $\{0,255\}$. $\widetilde x_{2,q}[k]$ is quantized to $0$ if $\widetilde x_{2,q}[k]\leq 128$ and $255$ if $\widetilde x_{2,q}[k]> 128$. One-bit information can be sent to AP1 for each $\widetilde x_{2,q,m}[k]$ when $m=1$. 

Based on ($11$), given $\widetilde x_{2,q,m}[k]$, a reconstructed soft bit $\hat x_{2,q}[k]$ at AP1 is given by 
\begin{align}
\hat x_{2,q}[k]  \buildrel \Delta \over = \frac{{ {\widetilde x_2[k]}}}{{|{h_{2,\max }}{|^2}}}= \frac{{\left( {\frac{{{{\widetilde x}_{2,q,m}}[k]}}{{255}} - \beta} \right)}}{\alpha }.
\end{align}
Notice that $|{h_{2, \max }}{|^2}$ is not available at AP1. Therefore, we combine and quantize the information from AP1 and AP2 as follows, assuming $|{h_{1, \max }}{|^2}=|{h_{2, \max }}{|^2}$,
\begin{align}
&\widetilde {x}_{q}^{\rm{co}}[k] = \left(\left(\frac{{ {\widetilde x_1[k]}}}{{|{h_{1,\max }}{|^2}}}+\hat x_{2,q}[k]\right)~\alpha'  + \beta\right) \times 255 \notag \\
&= \left(\left(\frac{{ {{y_1}[k] \cdot {h_1}[k]}}}{{|{h_{1,\max }}{|^2}}}+\frac{{\left( {\frac{{{{\widetilde x}_{2,q,m}}[k]}}{{255}} - \beta} \right)}}{\alpha }\right)~\alpha'  + \beta\right) \times 255.
\end{align}
As $\alpha$ does in (11), the parameter $\alpha'$ is used to control the range of $\widetilde {x}_{q}^{\rm{co}}[k]$, i.e., $\widetilde {x}_{q}^{\rm{co}}[k]$ falls in between $0$ and $255$. In our experiments, $\alpha'=\alpha/2$ achieves good decoding performance. The quantized soft bits ${ \widetilde {X}_{q}^{\rm{co}}=( \widetilde {x}_{q}^{\rm{co}}[1],..., \widetilde {x}_{q}^{\rm{co}}[k],...)}$ are fed into the Viterbi decoder at AP1 to decode $C^i$. In the next section, we experimentally study the effect of $m$ on the average AoI.

\begin{figure}[t]
\centerline{\includegraphics[trim=0 0 0 00,width=0.5\textwidth]{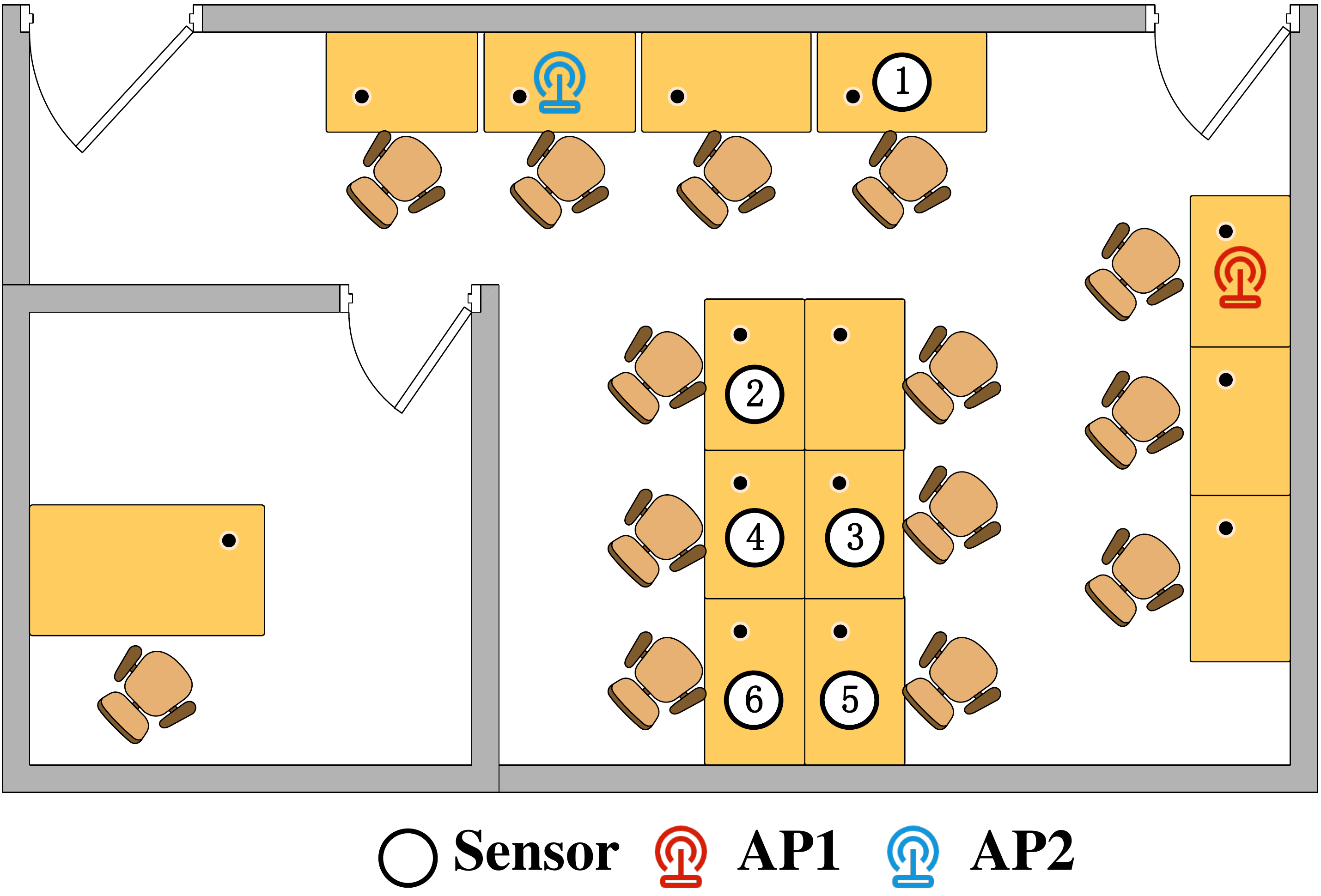}}
\caption{Layout of an indoor status update system used in our experiments. The positions of the two APs are fixed, and the sensors are distributed in different positions. In addition, we vary the transmit power of the sensor to get different SNR pairs $(SNR^{AP1}, SNR^{AP2})$.}
\label{exp_layout}
\end{figure}

\section{Experimental Evaluation} \label{sec:exp}
In this section, we present the experimental evaluation of the average AoI of status update systems with backbone-assisted cooperative APs. We first describe the PHY-layer implementation and the setup of our experiments, including the measurement of Ethernet backbone delay. After that, we present the experimental results of the average AoI performance in single-AP, Co-AP, and Soft-Co-AP.

\subsection{Implementation and Experimental Setup}\label{sec:exp1}
\subsubsection{PHY-layer Implementation and Experimental Setup} 
Our system prototypes are built on the USRP hardware \cite{Ettus} and the GNU Radio software \cite{G. FSF} with the UHD driver. We conduct experiments on our prototypes in an indoor laboratory with USRP devices acting as APs and sensors, as shown in Fig.~\ref{exp_layout}. 

At the PHY layer, all the systems adopt OFDM transmission. BPSK modulations and the $[133, 171]_8$ rate-1/2 convolutional codes defined in the IEEE 802.11 standards are used \cite{IEEE802.11ac}. Each update packet is preceded by a preamble before the payload. The preamble assists the receivers with packet synchronization and channel estimation. In our experiments, the duration of an update packet, $t_{pk}$, is around 1 microsecond (ms), which is calculated by 
\begin{align}
 {T_{pk}} = \left( 320 + \frac{{768\times 8\times 2}}{{48}} \times 80 \right) / ({2\times10^{7}})~s  \approx 1~ms,
\end{align}
where $320$ is the number of samples in the preamble. $80$ is the total number of samples in one OFDM symbol, consisting of a $64$-FFT OFDM symbol and a $16$-sample cyclic prefix. Although an OFDM symbol has $64$ subcarriers, only $48$ subcarriers are used for data transmission. $768\times 8\times 2 /48$ gives the number of OFDM symbols in an update packet with $768$ bytes and a channel coding rate of $1/2$. $2\times10^{7}$ means the bandwidth is $20$MHz. To facilitate packet decoding, we add a guard interval in each time slot so that the total duration of a time slot is $1.2$ms. 
\begin{figure}[t]
\centerline{\includegraphics[width=0.5\textwidth]{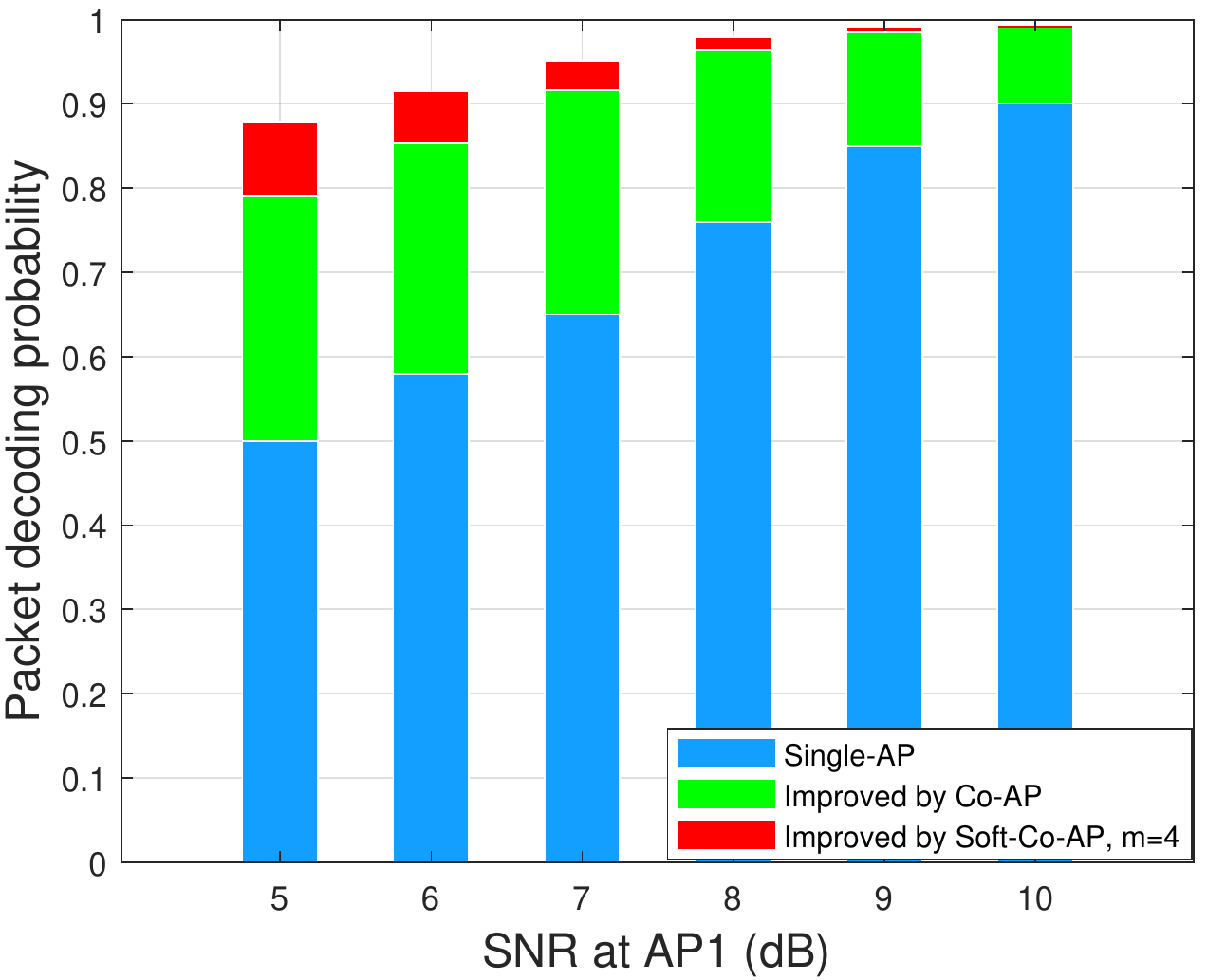}}
\caption{The PHY-layer packet decoding probability. We vary the SNR at AP1 from $5$dB to $10$dB, and the SNR at AP2 is larger than that at AP1 by $1$dB. The blue bars represent the packet decoding probability of single-AP, the green bars represent the improvement of the packet decoding probability by Co-AP, and the red bars represent the improvement of the packet decoding probability by Soft-Co-AP when $m=4$.}  
\label{exp_decoding_probability}
\end{figure}

In our experiments, we perform controlled experiments for different received SNRs from $5$dB to $10$dB. Specifically, we place USRPs (sensors) at different locations (see Fig.~\ref{exp_layout}) and control the transmit power of USRPs to simulate different received SNR pairs $(SNR^{AP1}, SNR^{AP2})$ at the two APs.  We use AP1 to schedule the transmission of $50,000$ packets of a sensor for each SNR pair $(SNR^{AP1}, SNR^{AP2})$. To compute the average AoI, we first collect the decoding outcomes (i.e., with and without joint decoding) of the sensors at the APs under different SNR pairs $(SNR^{AP1}, SNR^{AP2})$. Since sensors are independent in TDMA, after collecting the decoding outcomes at different positions, we can generate traces to drive the AoI simulations with a large number of sensors. Based on the successful or failed decoding in each TDMA round, we generate traces to drive the AoI simulations for different systems. In particular, the instantaneous AoI can be collected in each time slot, from which we compute the average AoI.

\begin{figure}[t]
\centerline{\includegraphics[width=0.5\textwidth]{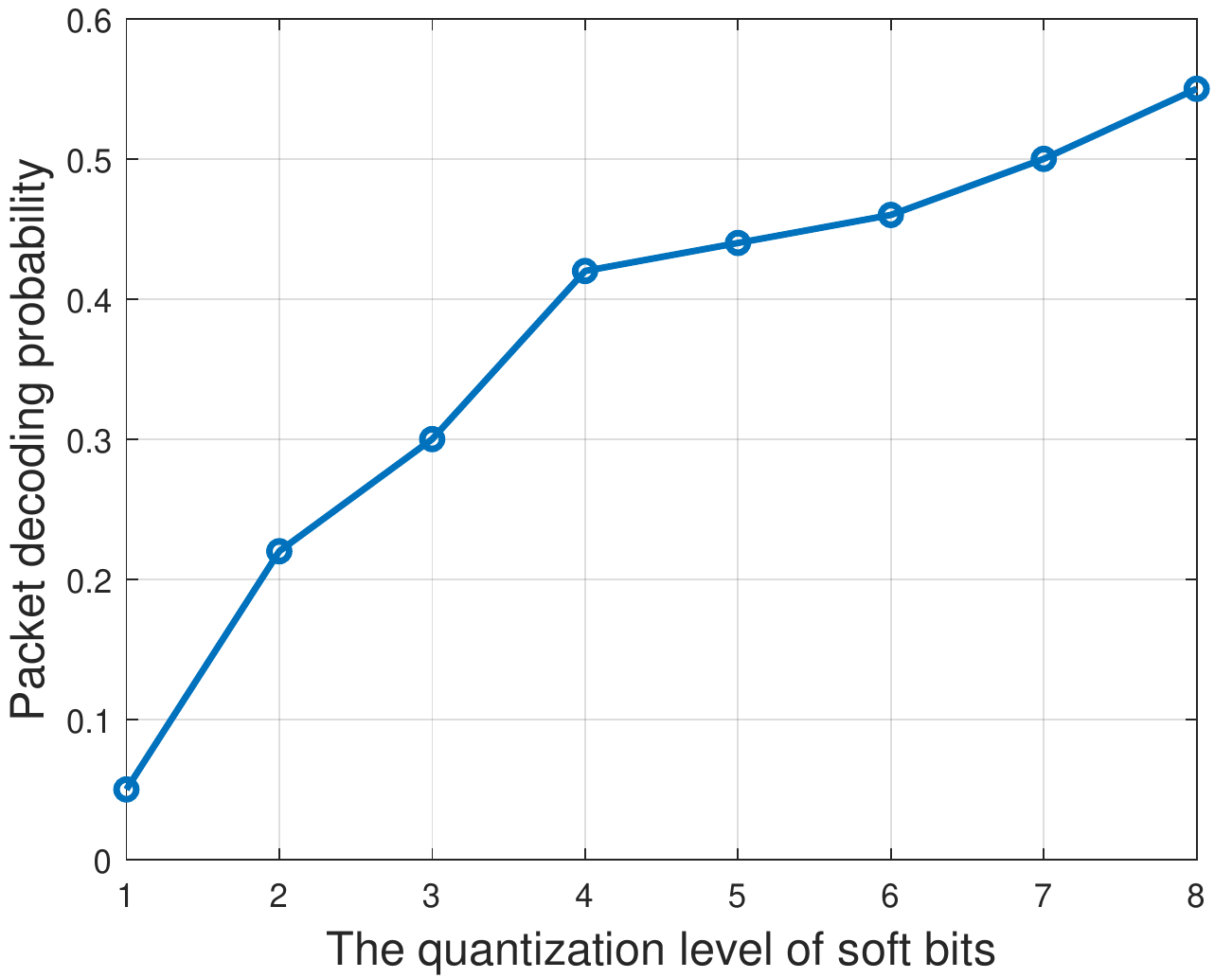}}
\caption{The PHY-layer packet decoding probability for diversity combining using different quantized levels of soft bits, where $SNR_i^{AP1}=5$dB and $SNR_i^{AP2}=6$dB. }  
\label{exp_quantization_bits}
\end{figure}

\subsubsection{Backbone Delay Measurement} 
We use the Linux PING command to measure backbone delay. PING is a network software that measures the round-trip time of messages (called PING packets) sent from the originating host to the destination host, which are echoed back to the source.\footnote{We note that the cooperative APs in the Ethernet backbone typically work like switches that forward trafﬁc based on MAC addresses. PING adds additional delays because it uses the ICMP protocol. ICMP is part of Internet Protocol related to the network layer \cite{A. Leon-Garcia}. If delays are directly measured between switches, we need to get into APs and change the lower-layer code (related to Ethernet). This is not feasible with many commercial APs which do not open up their codes (i.e., they are not open source). However, compared with the queuing and processing delays, the additional delays introduced by PING at the network layer are relatively small \cite{H. Pan4}.} Specifically, we adopt the experimental setup with hierarchical switches and wireless devices presented in \cite{H. Pan4} for measuring the packet forwarding delay between APs, where we purposely generate high trafﬁc to simulate a challenging WLAN-Ethernet environment. Moreover, we note
that PING measures the round-trip delay. As in \cite{H. Pan4}, symmetric and bidirectional trafﬁc ﬂows are established so that we can further assume that the one-way backbone delay is half of the round-trip delay of PING. We refer interested readers to Appendix \ref{appendix1} and \cite{H. Pan4} for the detailed setup.

We next discuss the payload of PING packets used to measure backbone delay. Recall that in Co-AP, AP2 needs to forward a decoded packet to AP1. To do so, the wireless packets destined for AP1, but received at AP2, are encapsulated in Ethernet frames for forwarding to AP1 (i.e., this procedure is referred to as Ethernet Tunneling in \cite{H. Pan4}). Hence, metadata such as the MAC header of a wireless packet should be included in the encapsulated Ethernet frame, e.g., AP2 can examine the MAC address in the MAC header to ﬁnd out that the received packets are destined for AP1. In our implementation, we assume that each wireless uplink packet has a 30-bytes MAC header and 768-bytes data so that the payload of a PING packet is $798$ bytes when a decoded packet needs to be forwarded to AP1 (i.e., the PING size is $798$ bytes when executing the PING command).

For Soft-Co-AP, AP2 needs to forward the quantized soft bits of a coded packet to AP1, where each soft bit is represented by $m$ quantization bits. Since we use the rate-1/2 convolutional codes, the size of the quantized soft bits for each coded packet is $1536m(=2\times m\times 768)$ bytes. Thus, to measure the backbone delay when AP2 forwards the quantized soft bits to AP1, we set the payload of PING size to $(1536m+30)$ bytes for different $m$. Furthermore, notice that the maximum transmission unit (MTU) of an Ethernet frame is usually $1500$ bytes. When the PING size exceeds $1500$ bytes, the whole payload will be separated and encapsulated into multiple PING packets. For example, a PING size of $798$ bytes requires only one PING packet; when the PING size is $12318$ bytes (i.e., $m=8$), nine PING packets are required. The more PING packets, the more backbone delay is required, as will be presented in the following subsection.

\begin{figure}[t]
\centerline{\includegraphics[width=0.5\textwidth]{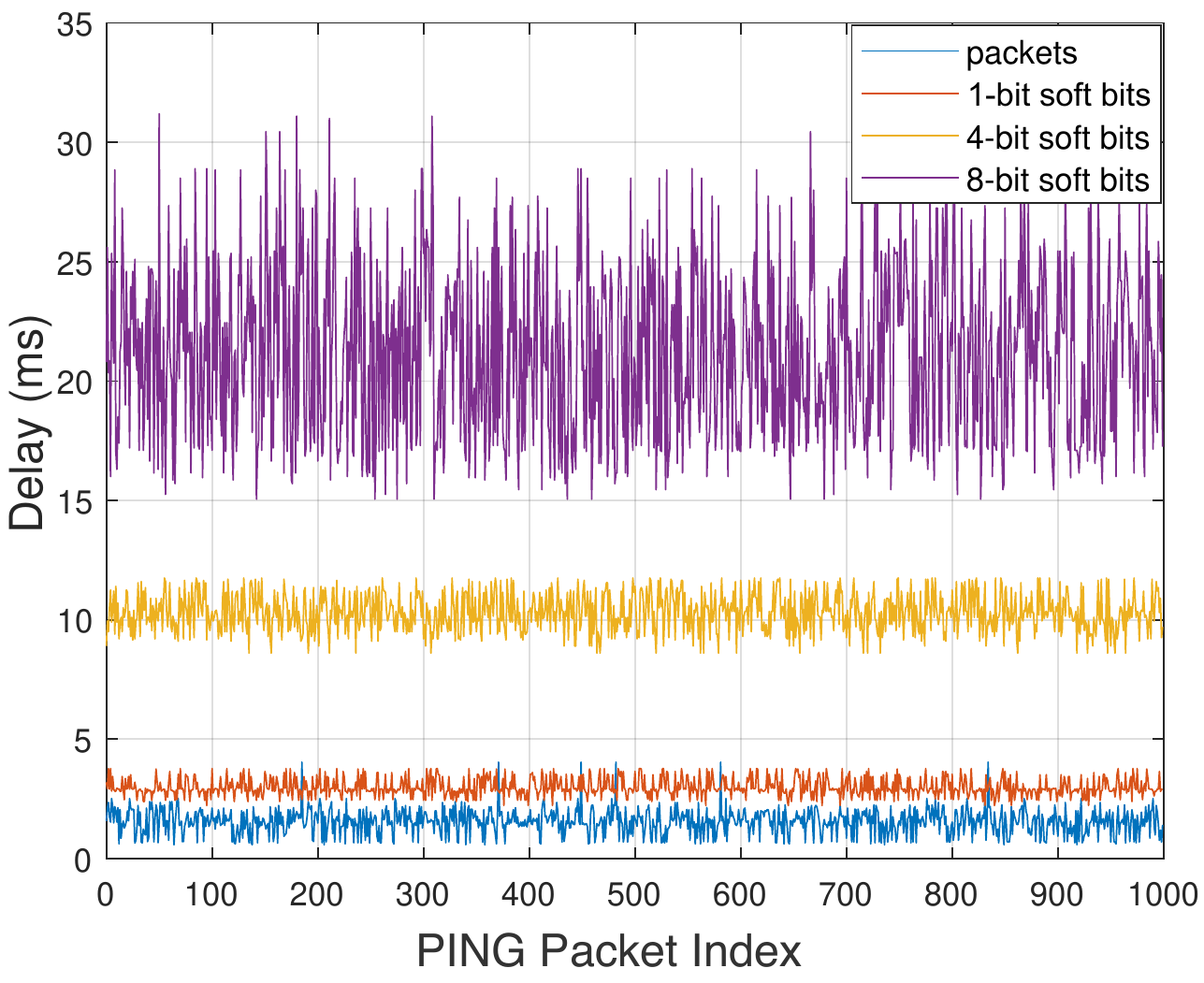}}
\caption{The measured real backbone delay samples.}  
\label{exp_backbone_delay}
\end{figure}

\subsection{Experimental Results}\label{sec:exp2}
In Section \ref{sec:decoded_packcet}, we have previewed the average AoI of single-AP and Co-AP systems when the difference in SNR between AP1 and AP2 varies (see Fig.~\ref{exp_co_ap}). Now, this subsection further presents the PHY-layer packet decoding probability obtained in our SDR prototype and the backbone delay statistics. After that, the average AoI performances of single-AP, Co-AP, and Soft-Co-AP are presented and compared. Unless specified otherwise, we consider an SNR-balanced scenario, in which all sensors have the same received SNR at AP1 or AP2. We vary the SNR of sensors at AP1 from $5$dB to $10$dB. The SNR at AP1 is smaller than that at AP2 by $1$dB to simulate different signal paths.

\subsubsection{PHY-layer Packet Decoding Probability}
Let us look at the PHY-layer decoding probability plotted in Fig.~\ref{exp_decoding_probability}. We examine the packet decoding probabilities at the primary AP (AP1) for different systems. The blue bars represent the packet decoding probability of single-AP, and the green bars represent the improvement of Co-AP in packet decoding probability over single-AP. We see that the improvement in packet decoding probability is significant when Co-AP is used. For example, when $SNR^{AP1}=5$dB and $SNR^{AP2}=6$dB, Co-AP increases the packet decoding 
probability by around $30$\% over single-AP.

Soft-Co-AP further forwards the quantized soft bits of coded packets to AP1, when AP2 cannot decode update packets. The red bars in Fig.~\ref{exp_decoding_probability} represent the further improvement of packet decoding probability by Soft-Co-AP. Since soft bits can be represented by different $m$, Fig.~\ref{exp_decoding_probability} plots the case when $m=4$. Fig.~\ref{exp_quantization_bits} plots the packet decoding probability versus $m$, when $SNR^{AP1}=5$dB and $SNR^{AP2}=6$dB (other SNR pairs $(SNR^{AP1}, SNR^{AP2})$ lead to similar observations). From Fig.~\ref{exp_quantization_bits}, we see that a larger $m$ leads to a better joint packet decoding performance. However, as mentioned earlier, a larger $m$ also leads to a larger backbone delay.  We next present the backbone delay statistics measured in the experiments.

\begin{figure}[t]
\centerline{\includegraphics[width=0.5\textwidth]{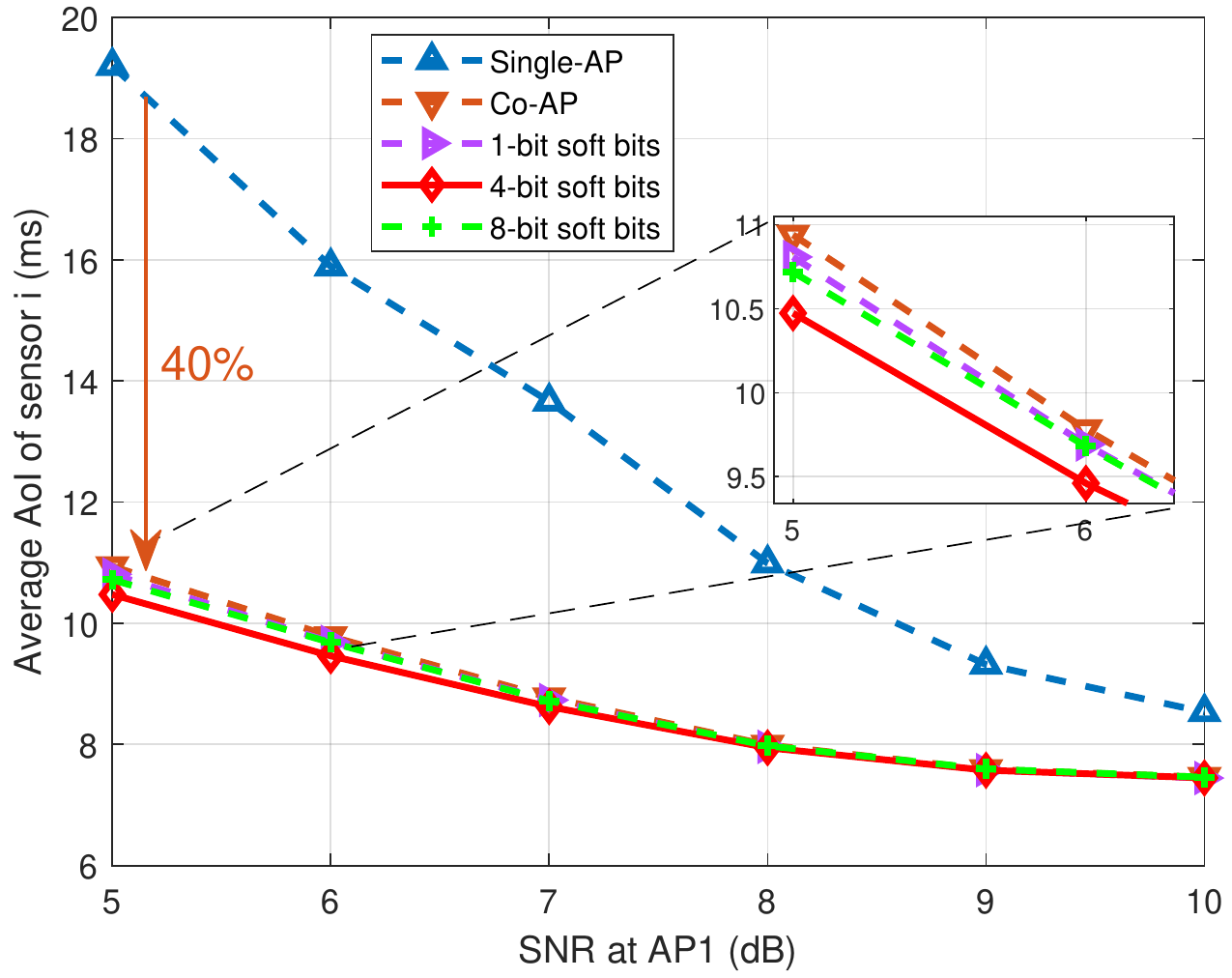}}
\caption{The average AoI performance of the system for Single-AP, Co-AP, and Soft-Co-AP when all sensors have the same received SNR at the APs. The number $N$ of sensors is set to $10$, and the SNR of sensors at AP2 is larger than that at AP1 by $1$dB.}  
\label{exp_aoi_snr}
\end{figure}

\subsubsection{Backbone Delay Statistics}\label{exp2.2}
We now look at the backbone delay statistics. Fig.~\ref{exp_backbone_delay} plots the backbone delay samples when AP2 forwards different types of data in our experiments, namely the decoded packets in the Co-AP system and the quantized soft bits ($m\in\{1,4,8\}$) in the Soft-Co-AP system. From Fig.~\ref{exp_backbone_delay}, we see that the backbone delays are dynamically changing, e.g., in Soft-Co-AP, the instantaneous backbone delay fluctuates between $15$ms and $30$ms when $m=8$. Furthermore, the more PING packets required (i.e., a larger $m$), the more time for AP1 to receive the information from AP2. For example, the average backbone delay is around $10$ms and $20$ms, when $m=4$ and $m=8$, respectively. Hence, although a larger $m$ results in a higher decoding probability, it also leads to a larger average backbone delay. Since both the decoding probability and the backbone delay affect the information freshness, the average AoI performance should be carefully studied, as presented next. 

\begin{figure*}[t]
\centering  
\subfigure[The SNR of all sensors is $5$dB at AP1 and $6$dB at AP2.]{
\label{exp_aoi_sensors_samesnr}
\includegraphics[width=0.5\textwidth]{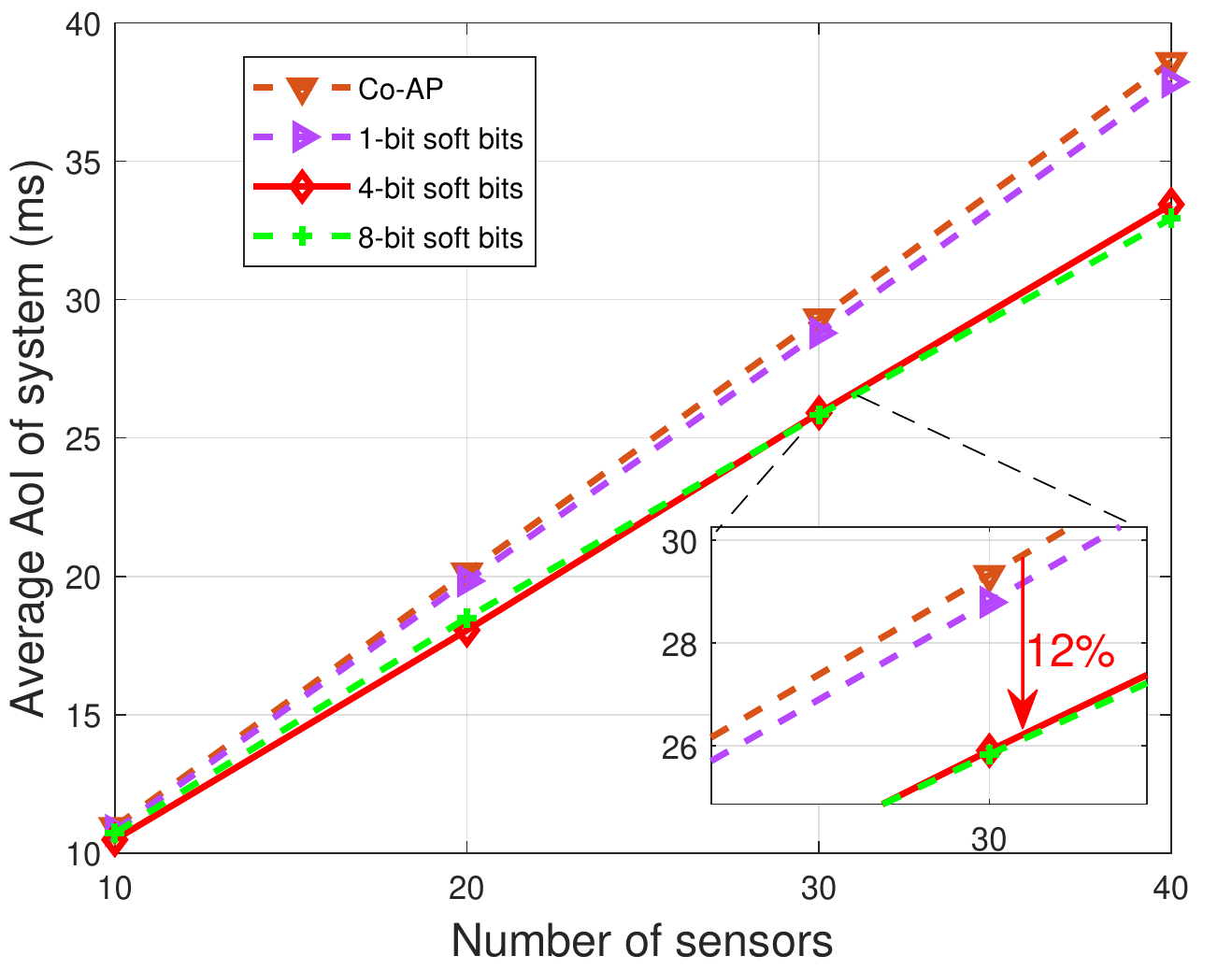}}\subfigure[Sensors have different SNRs at AP.]{
\label{exp_aoi_sensors_diffsnr}
\includegraphics[width=0.5\textwidth]{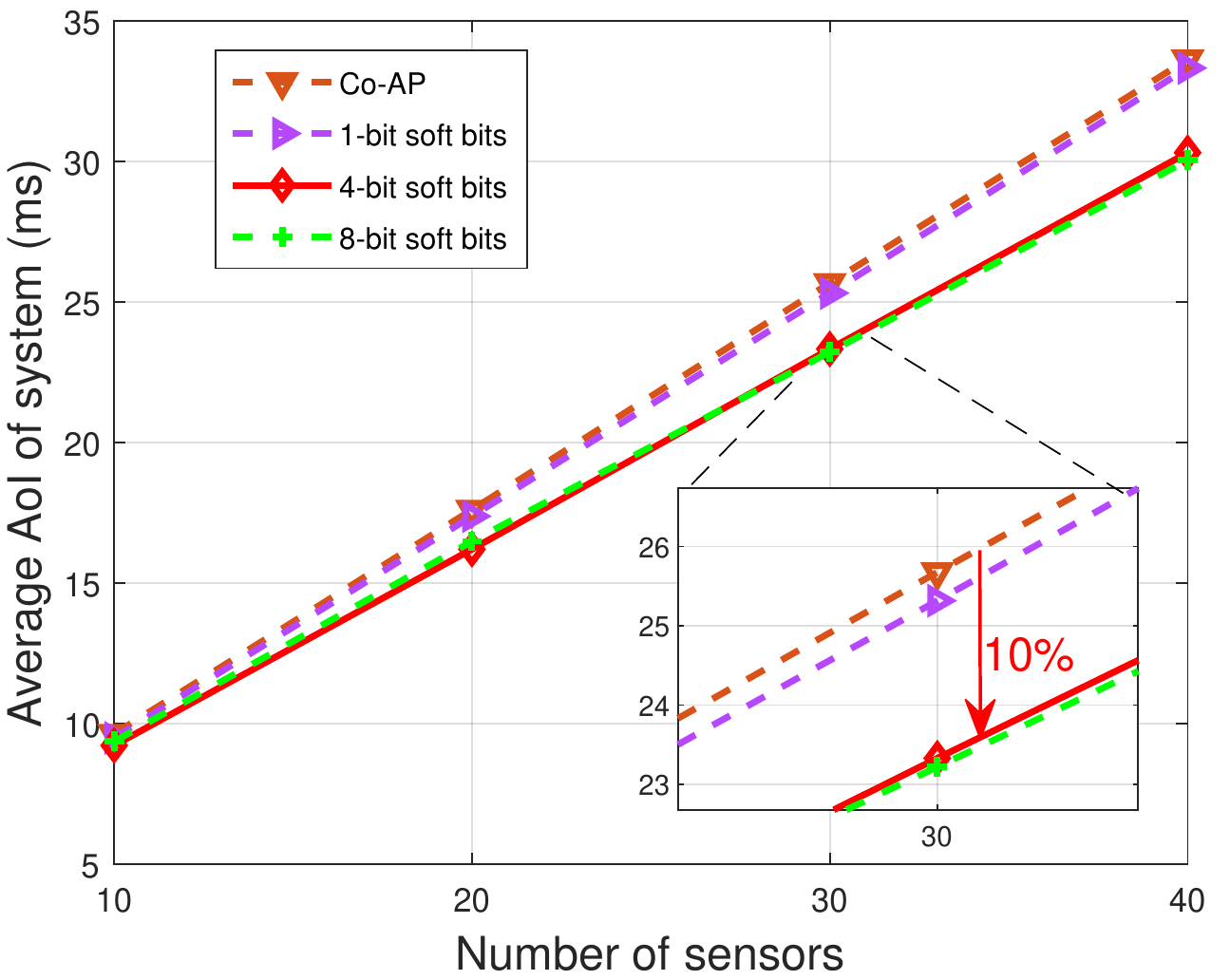}}
\caption{The average AoI performance for Co-AP and Soft-Co-AP systems, versus the number $N$ of sensors. We consider the SNR-balanced scenario with $SNR^{AP1}=5$dB and $SNR^{AP2}=6$dB.}
\label{exp_aoi_sensors}
\end{figure*}

\subsubsection{Average AoI Comparison}\label{exp3}
First, let us look at the average AoI performance of different systems in the SNR-balanced scenario. We set the number of sensors to $10$. The average AoI of a sensor (say sensor $i$) under different systems versus the SNR at AP1 is shown in Fig.~\ref{exp_aoi_snr} (the SNR at AP2 is larger than that at AP1 by $1$dB). We can see that Co-AP has a much lower average AoI than single-AP, thanks to the decoded packets forwarded by AP2 through the backbone. For example, when $SNR^{AP1}=5$dB, Co-AP reduces the average AoI of single-AP by around $40$\%. This is consistent with the preliminary experimental results shown in Fig.~\ref{exp_co_ap}. 

As for the average AoI of Soft-Co-AP, Fig.~\ref{exp_aoi_snr} shows that when the SNR is low (i.e., $5$dB$\sim7$dB), Soft-Co-AP further reduces the average AoI compared with Co-AP. When the SNRs at AP1 are larger than $7$dB, the improvement by Soft-Co-AP is small because Co-AP can already perform well, i.e., AP2 can decode update packets so there is no need to forward soft bits. 

Interestingly, as indicated by Fig.~\ref{exp_aoi_snr}, the average AoI performance of $m=4$ is slightly better than that of $m=8$, even though $m=8$ leads to a higher packet decoding probability. When the number of sensors is $10$, the total duration of a round is $12ms$. However, the average backbone delay when $m=8$ is around $20ms$, as shown in Fig.~\ref{exp_backbone_delay}. This means that even if AP1 can receive the soft bits from AP2 after $20ms$ and the ``old'' update packet can be recovered successfully at AP1, a ``new'' update packet has been sent in a new TDMA round. Therefore, the new update packet may be successfully received by AP1 before the old update packet, i.e., the recovery of the old update packet does not help reduce the AoI. On the other hand, even if no new packets can be received, the instantaneous AoI is still large when the old packet is decoded due to the considerable backbone delay. By contrast, the average backbone delay when $m=4$ is around $10ms$, which is smaller than a TDMA round. In other words, AP1 can try to decode the old packet again before a new update packet is sent in a new TDMA round. If the old packet can be recovered, the average AoI can be reduced, as indicated in Fig.~\ref{aoi_examples} of Section \ref{sec:decoded_packcet}. 

Now, we further study the scenarios with more sensors and the average AoI of the whole network in Fig.~\ref{exp_aoi_sensors_samesnr}. We consider the SNR-balanced scenario with $SNR^{AP1}=5$dB and $SNR^{AP2}=6$dB and vary the number of sensors. We find that the larger the number of sensors, the more significant the improvement of Soft-Co-AP over Co-AP. For example, when $N=30$, Soft-Co-AP reduces the average AoI by $12$\% over Co-AP. With the increase in the number of sensors, the time of one TDMA round increases accordingly. In this case, if a sensor fails to update in a TDMA round, it needs to wait a longer time for the next update opportunity, i.e., the time interval between two consecutive updates becomes larger. With a longer TDMA round, in most scenarios, AP1 can try to decode the old packet again before a new TDMA round starts, even when $m=8$. Hence, the higher packet decoding probability in Soft-Co-AP can effectively reduce the average AoI. In addition, since the duration of a TDMA round is large, $m=4$ and $m=8$ lead to almost the same average AoI performance, indicating that $m=4$ is a more viable option since it reduces the backbone traffic.

\textbf{SNR-imbalanced scenario:} In addition, we look at the average AoI performance of Soft-Co-AP in an SNR-imbalanced scenario, as shown in Fig.~\ref{exp_aoi_sensors_diffsnr}. Specifically, sensors have different received SNRs at AP1 or AP2. We assume that half of the sensors have an SNR of $5$dB at AP1 and $6$dB at AP2, and the other half have an SNR of $10$dB at AP1 and $11$dB at AP2. In Fig.~\ref{exp_aoi_sensors_diffsnr}, we can observe the same phenomenon in Fig.~\ref{exp_aoi_sensors_samesnr}, i.e., the larger the number of sensors, the more significant the improvement of Soft-Co-AP over Co-AP. It is enough to adopt the number of quantization bits per soft bit that is neither too large nor too small (i.e., $m=4$) to achieve a good average AoI performance. Therefore, Soft-Co-AP is a practical solution to improve information freshness with a large number of sensors. 
 
\begin{figure}[t]
\centerline{\includegraphics[width=0.5\textwidth]{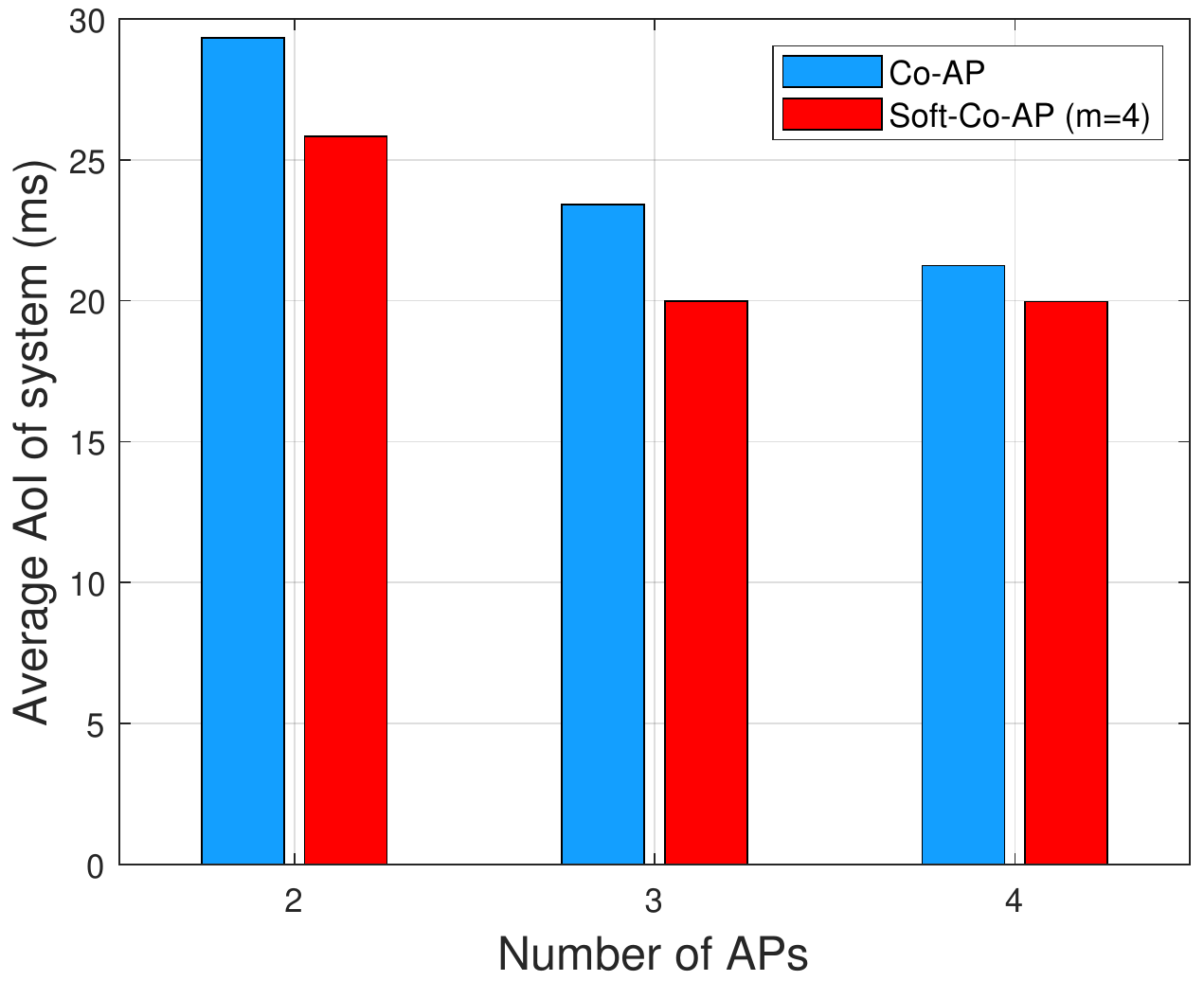}}
\caption{The average AoI performance versus the number of APs. The number $N$ of sensors is set to $30$, and the SNR of each sensor is $5$dB at AP1 and $6$dB at other secondary APs.}  
\label{exp_multiple_ap}
\end{figure} 

\textbf{More than two cooperative APs:} In practice, it is likely that more than two APs can receive wireless signals from the same sensor. Here, we further consider an experiment with more than two cooperative APs. The average AoI of the network under different numbers of cooperative APs is  shown in Fig.~\ref{exp_multiple_ap}. When there are more than two cooperative APs, the experimental setup is the same as the two-AP scenarios, except that one AP serves as the primary AP and all the other APs serve as the secondary APs. In Fig.~\ref{exp_multiple_ap}, the number of sensors $N$ is set to $30$, and the SNR of each sensor is $5$dB at the primary AP and $6$dB at the secondary APs. As shown in Fig.~\ref{exp_multiple_ap}, the average AoI is reduced for both Co-AP and Soft-Co-AP when there are three cooperative APs, compared with the case with two cooperative APs only. This indicates that more APs increase spatial diversity such that an update packet can have a higher chance of being received, thus improving information freshness.

Moreover, we see in Fig.~\ref{exp_multiple_ap} that when the number of cooperative APs increases from three to four, the average AoI does not reduce further. This is because, in the three-AP scenario, the average AoI almost reaches the optimal average AoI in a TDMA single-AP system. It is easy to figure out that the optimal average AoI of a TDMA system with $30$ sensors is $19.2ms$, where each time slot is of $1.2ms$ duration and each update packet is received successfully by the AP. The average AoI of the three-AP and four-AP systems is slightly higher than the optimal average AoI of $19.2ms$ because an additional backbone delay is induced when cooperative APs forward information over the backbone network.   

\section{Conclusions}\label{sec:conl}
We have demonstrated a viable solution for timely status update systems using backbone-assisted cooperative APs. This paper is the first attempt to study backbone-assisted cooperative APs to improve information freshness. We first investigate Co-AP, a system where the secondary AP forwards only the decoded packets to the primary AP through the backbone. Thanks to the forwarded decoded packets, the update packets that fail to be decoded by the primary AP can be successfully recovered, thus reducing the average AoI significantly compared with the traditional single-AP system. 

We further investigate an improved Co-AP system, referred to as the Soft-Co-AP system.  Soft-Co-AP can forward the soft bits of a coded packet to the primary AP through the backbone when the secondary AP fails to decode the packet. Then the primary AP tries to decode the packet again with the help of soft bits. An interesting question here is the tradeoff between joint decoding probability and backbone delay under different quantization bits per soft bit, both of which affect the average AoI. We experimentally explore the impact of the number of quantization bits per soft bit on the average AoI in Soft-Co-AP.

Experimental results on our software-defined radio prototype indicate that Soft-Co-AP further improves the average AoI performance over Co-AP, especially when the number of sensors in the TDMA network is large. Due to the tradeoff between joint decoding performance and backbone delay, the number of quantization bits per soft bit is usually neither too large nor too small to achieve a good average AoI performance. 

Overall, Soft-Co-AP is a practical solution to improve the information freshness of a large number of IoT devices. Moving forward, we plan to extend the current TDMA scheme to advanced non-orthogonal multiple access schemes to further improve the AoI performance \cite{H. Pan1}. Another possible direction is to incorporate Co-AP/Soft-Co-AP in random access scenarios, such as a real Wi-Fi system with carrier sensing and collision avoidance capabilities \cite{IEEE802.11ac}. We believe that the backbone-assisted solution applies to these scenarios, and their performance improvements are worthy of future investigation.

\begin{figure}[t]
\centerline{\includegraphics[width=0.5\textwidth]{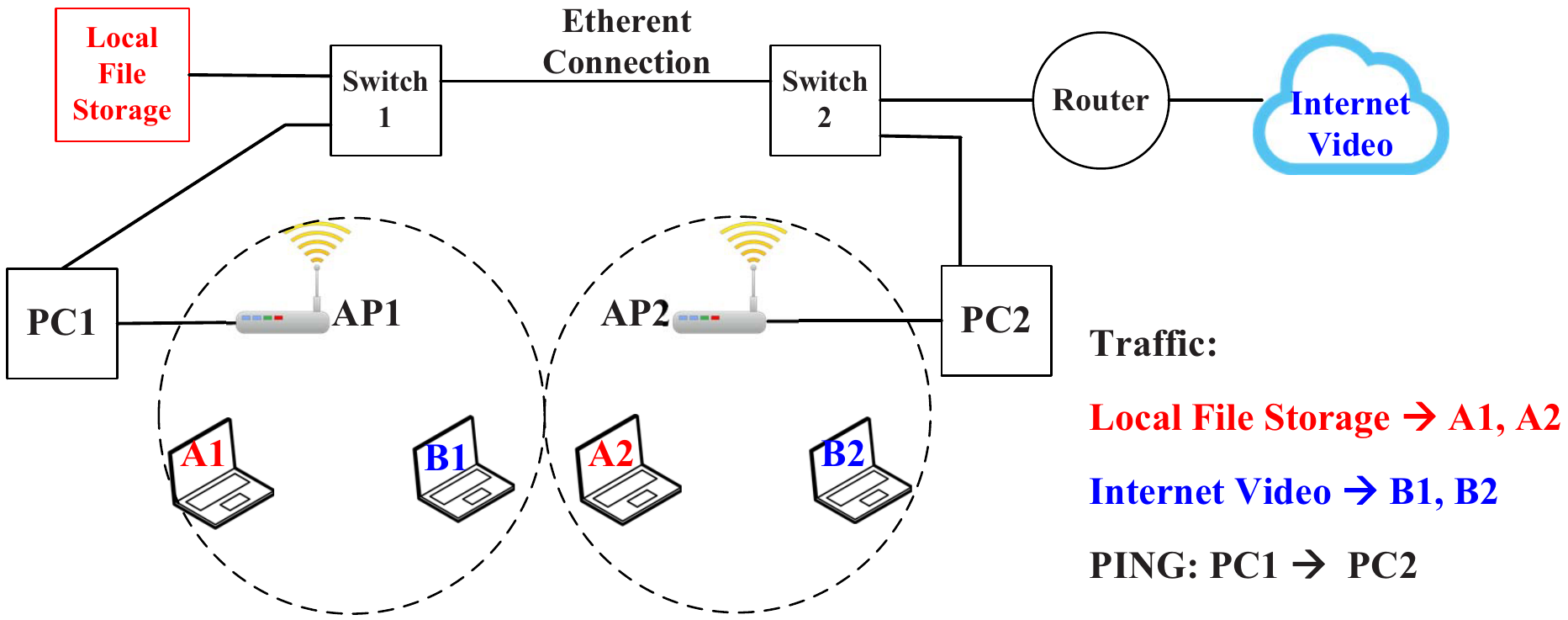}}
\caption{Experimental Setup for the Backbone Delay Measurement \cite{H. Pan4}.}  
\label{exp_delay_measure}
\end{figure} 

\appendices
\section{Experimental Setup for the Backbone Delay Measurement} \label{appendix1}
The experimental setup for our backbone delay measurements follows \cite{H. Pan4}. As shown in Fig. \ref{exp_delay_measure}, two switches are interconnected via Ethernet, and a personal computer (PC) is connected to each switch. Each PC is equipped with a wireless network interface card to serve as an AP. The two PCs (also APs) connected by two switches simulate a practical scenario where packets from the secondary AP may reach the primary AP through more than one switch (i.e., a hierarchy structure). As shown in Fig. \ref{exp_delay_measure}, AP1 serves two users (A1 and B1), and AP2 serves two users (A2 and B2). In addition, local file storage is connected to switch 1, and a router is connected to switch 2.

We measure the backbone delay by PC1 PING PC2. We deliberately generate high traffic in the current experimental setup to simulate a challenging WLAN-Ethernet environment. Specifically, users A1 and A2 download a large number of files from the local file storage. Meanwhile, users B1 and B2 receive video streams from the Internet. By doing so, PC1's PING packets ``compete'' with the traffic from the file storage to user A2 at switch 1, and also ``compete'' with the traffic from the Internet to user B2 at switch 2 in the backbone network. This traffic setup simulates the queuing delay of each switch. In addition, with such symmetric and bidirectional traffic flows, we can assume that the one-way backbone delay is half of the round-trip delay of PING.



\begin{thebibliography}{00}
\bibitem{A. Zanella} A. Zanella, N. Bui, A. Castellani, L. Vangelista, and M. Zorzi, ``Internet of Things for smart cities,'' \emph{IEEE Internet Things J.}, vol. 1, no. 1, pp. 22--32, Feb. 2014.

\bibitem{M. A. Abd-Elmagid} M. A. Abd-Elmagid, N. Pappas, and H. S. Dhillon, ``On the role of age of information in the Internet of Things,'' \emph{IEEE Commun. Mag.}, vol. 57, no. 12, pp. 72--77, Dec. 2019.

\bibitem{D. Ciuonzo} D. Ciuonzo, P. S. Rossi, and P. K. Varshney, ``Distributed detection in wireless sensor networks under multiplicative fading via generalized score tests,'' \emph{IEEE Internet Things J.}, vol. 8, no. 11, pp. 9059--9071, Jun. 2021.

\bibitem{M. A. Al-Jarrah} M. A. Al-Jarrah, M. A. Yaseen, A. Al-Dweik, O. A. Dobre, and E. Alsusa, ``Decision fusion for IoT-based wireless sensor networks,'' \emph{IEEE Internet Things J.}, vol. 7, no. 2, pp. 1313--1326, Feb. 2020.

\bibitem{X. Cheng} X. Cheng, D. Ciuonzo, P. S. Rossi, X. Wang, and W. Wang, ``Multi-bit \& sequential decentralized detection of a noncooperative moving target through a generalized Rao test,'' \emph{IEEE Trans. Signal Inf. Process. Netw.}, vol. 7, pp. 740--753, Nov. 2021.

\bibitem{H. Zhou} H. Zhou, W. Xu, J. Chen, and W. Wang, ``Evolutionary V2X technologies toward the Internet of vehicles: Challenges and opportunities,'' \emph{Proc. IEEE}, vol. 108, no. 2, pp. 308--323, Feb. 2020.

\bibitem{S. Kaul}S. Kaul, M. Gruteser, V. Rai, and J. Kenney, ``Minimizing age of information in vehicular networks,'' \emph{Proc. IEEE SECON}, 2011, pp. 350--358.

\bibitem{A. Kosta} A. Kosta, N. Pappas, and V. Angelakis, ``Age of information: A new concept, metric, and tool,'' \emph{Found. Trends Netw.}, vol. 12, no. 3, pp. 162--259, 2017.



\bibitem{X. Chen} X. Chen and S. S. Bidokhti, ``Benefits of coding on age of information in broadcast networks,'' \emph{IEEE ITW}, pp. 1--5, Aug. 2019.

\bibitem{H. Pan2} H. Pan, S. C. Liew, J. Liang, V. C. M. Leung, and J. Li, ``Coding of multi-source information streams with age of information requirements,''\emph{IEEE J. Sel. Areas Commun.}, vol. 39, no. 5, pp. 1427--1440, May 2021.

\bibitem{Munari} A. Munari, ``Modern random access: An age of information perspective on irregular repetition slotted ALOHA,'' \emph{IEEE Trans. Comm.}, vol. 69, no. 6, pp. 3572--3585, Jun. 2021.

\bibitem{H. Pan1} H. Pan, J. Liang, S. C. Liew, V. C. M. Leung, and J. Li, ``Timely information update with nonorthogonal multiple access,'' \emph{IEEE Trans. Ind. Informat.}, vol. 17, no. 6, pp. 4096--4106, Jun. 2021.

\bibitem{Grybosi} J. F. Grybosi, J. L. Rebelatto, and G. L. Moritz, ``Age of information of SIC-aided massive IoT networks with random access'' \emph{IEEE Internet Things J.}, vol. 9, no. 1, pp. 662--670, Jan. 2022.

\bibitem{A. Goldsmith}A. Goldsmith, \textit{et al.}, \emph{Wireless Communications}, Cambridge University Press, 2005.

\bibitem{ShuCoding} S. Lin and D. J. Costello, \emph{Error Control Coding (2nd Edition)}, Prentice Hall, Jun. 2004.

\bibitem{Spiral Project} Spiral Project, ``Viterbi decoder software generator,'' www.spiral.net/software/viterbi.html. 

\bibitem{X. Wang} X. Wang, C. Chen, J. He, S. Zhu, and X. Guan, ``AoI-aware control and communication co-design for industrial IoT systems,''\emph{IEEE Internet Things J.}, vol. 8, no. 10, pp. 8464--8473, May 2021.


\bibitem{L. Guo} L. Guo, Z. Chen, D. Zhang, K. Liu, and J. Pan, ``Age-of-information-constrained transmission optimization for ECG-based body sensor networks,'' \emph{IEEE Internet of Things J.}, vol. 8, no. 5, pp. 3851--3863, Mar. 2021.

\bibitem{Z. Ling} Z. Ling, F. Hu, H. Zhang, and Z. Han, ``Age of information minimization in healthcare IoT using distributionally robust optimization,'' \emph{IEEE Internet of Things J.}, early access, Feb. 2022. doi: 10.1109/JIOT.2022.3150321.

\bibitem{Y. Li} Y. Li, Y. Xu, Q. Zhang, and Z. Yang, ``Age-driven spatially-temporally correlative updating in the satellite-integrated Internet of Things via Markov decision process,'' \emph{IEEE Internet Things J.}, early access, Jan. 2022. doi:10.1109/JIOT.2022.3142268.
 
\bibitem{Y. Inoue} Y. Inoue, H. Masuyama, T. Takine, and T. Tanaka, ``A general formula for the stationary distribution of the age of information and its
application to single-server queues,'' \emph{IEEE Trans. Inf. Theory}, vol. 65, no. 12, pp. 8305--8324, Dec. 2019.

\bibitem{A. M. Bedewy} A. M. Bedewy, Y. Sun, and N. B. Shroff, ``Minimizing the age of information through queues,'' \emph{IEEE Trans. Inf. Theory}, vol. 65, no. 8, pp. 5215--5232, Aug. 2019.

\bibitem{J. P. Champati} J. P. Champati, R. R. Avula, T. J. Oechtering, and J. Gross, ``Minimum achievable peak age of information under service preemptions and request delay,'' \emph{IEEE J. Sel. Areas Commun.}, vol. 39, no. 5, pp. 1365--1379, May 2021.

\bibitem{M. Xie} M. Xie, Q. Wang, J. Gong, and X. Ma, ``Age and energy analysis for LDPC coded status update with and without ARQ,'' \emph{IEEE Internet Things J.}, vol. 7, no. 10, pp. 10388--10400, Oct. 2020.

\bibitem{E. T. Ceran} E. T. Ceran, D. Gündüz, and A. György, ``Average age of information with hybrid ARQ under a resource constraint,'' \emph{IEEE Trans. Wireless Commun.}, vol. 18, no. 3, pp. 1900--1913, Mar. 2019.

\bibitem{M. Xie2} M. Xie, J. Gong, X. Jia, and X. Ma, ``Age and energy tradeoff for multicast networks with short packet transmissions,'' \emph{IEEE Trans. Commun.}, pp. 6106--6119, Sep. 2021.

\bibitem{H. Pan3} H. Pan, T.-T. Chan, V. C. M. Leung, and J. Li, ``Age of information in physical-layer network coding enabled two-way relay networks,'' \emph{IEEE Trans. Mob. Comput.}, early access, Apr. 2022. doi: 10.1109/TMC.2022.3166155.

\bibitem{Costa. M} M. Costa, M. Codreanu, and A. Ephremides, ``On the age of information in status update systems with packet management''. \emph{IEEE Trans. Inf. Theory}. vol. 62, no. 4, pp. 1897--1910, Apr. 2016.

\bibitem{S. Feng} S. Feng and J. Yang, ``Precoding and scheduling for AoI minimization in MIMO broadcast channels,'' \emph{IEEE Trans. Inf. Theory}, early access, Apr. 2022. doi: 10.1109/TIT.2022.3167618.

\bibitem{B. Yu} B. Yu and Y. Cai, ``Age of information in grant-free random access with massive MIMO,'' \emph{IEEE Wireless Commun. Lett.}, vol. 10, no. 7, pp. 1429--1433, Jul. 2021.

\bibitem{H. Pan4} H. Pan and S. C. Liew, ``Backbone-assisted wireless local area network,'' \emph{IEEE Trans. Mobile Comput.}, vol. 20, no. 3, pp. 830--845, Mar. 2021.

\bibitem{K. Tan} K. Tan, H. Liu, J. Fang, W. Wang, J. Zhang, M. Chen, and G. M. Voelker, ``SAM: Enabling practical spatial multiple access in wireless LAN,'' in \emph{Proc. ACM MobiCom}, 2009, pp. 49--60.

\bibitem{H. Balan} H. Balan, R. Rogalin, A. Michaloliakos, K. Psounis, and G. Caire, ``AirSync: Enabling distributed multiuser MIMO with full spatial multiplexing,'' \emph{IEEE/ACM Trans. Networking}, vol. 21, no. 6, pp. 1681--1695, Dec. 2013.

\bibitem{W. Zhou} W. Zhou, T. Bansal, P. Sinha, and K. Srinivasan, ``{BBN}: Throughput scaling in dense enterprise WLANs with bind beamforming and nulling,'' in \emph{Proc. ACM MobiCom}, 2014, pp. 165--176.

\bibitem{Irmer} R. Irmer \emph{et al.}, ``Coordinated multipoint: Concepts, performance, and field trial results,'' \emph{IEEE Commun. Mag.}, vol. 49, no. 2, pp. 102--111, Feb. 2011.

\bibitem{D. Lee} D. Lee \emph{et al.}, ``Coordinated multipoint transmission and reception in LTE-advanced: Deployment scenarios and operational challenges,'' \emph{IEEE Commun. Mag.}, vol. 50, no. 2, pp. 148--155, Feb. 2012.

\bibitem{Ammar} H. A. Ammar, R. Adve, S. Shahbazpanahi, G. Boudreau, and K. V. Srinivas, ``User-centric cell-free massive MIMO networks: A survey of opportunities, challenges and solutions,'' \emph{IEEE Commun. Surveys Tuts.}, vol. 24, no. 1, pp. 611--652, 1st Quart., 2022.

\bibitem{G. Woo} G. Woo, P. Kheradpour, D. Shen, and D. Katabi, ``Beyond the bits: Cooperative packet recovery using physical layer information,'' in \emph{Proc. ACM Mobicom}, 2007, pp. 147--158.

\bibitem{M. Gowda} M. Gowda, S. Sen, R. R. Choudhury, and S.-J. Lee, ``Cooperative packet recovery in enterprise WLANs,'' in \emph{Proc. IEEE INFOCOM}, 2013, pp. 1348--1356.

\bibitem{IEEE802.11ac}IEEE802.11ac-2013, ``Wireless LAN medium access control (MAC) and physical layer (PHY) specifications—amendment 4: Enhancements for very high throughput for operation in bands below 6GHz,'' in \emph{IEEE Std 802.11ac-2013}, pp.1--425, Dec. 2013.

\bibitem{G. FSF}G. FSF., ``GNU Radio - GNU FSF Project,'' Accessed: May 31, 2022. [Online]. Available: http://www.gnuradio.org/

\bibitem{Ettus}Ettus Inc., ``Universal software radio peripheral,'' Accessed: May 31, 2022. [Online]. Available: https://www.ettus.com/

\bibitem{L. Lu} L. Lu, L. You, and S. C. Liew, "Network-coded multiple access," \emph{IEEE Trans. Mob. Comput.}, vol. 13, no. 12, pp. 2853--2869, Dec. 2014.

\bibitem{A. Leon-Garcia} A. Leon-Garcia, \emph{Communication Networks: Fundamental Concepts and Key Architectures}. 2nd Ed., New York, NY, USA: McGraw-Hill, 2003.


\end{thebibliography}
\end{document}